\documentclass[notitlepage,11pt]{article}

\newcommand{\Pb}{\hbox{{I}\kern-.1667em\hbox{P}}}

\usepackage{graphicx}
\usepackage{amsmath}
\usepackage{amssymb}
\usepackage[margin = 1in]{geometry}
 \usepackage{amsthm}
\usepackage{lscape}
\usepackage[final]{changes}
\definechangesauthor[color=red]{TRH}
\definechangesauthor[color=blue]{DOW}
\usepackage{float}

\usepackage{arydshln}
\usepackage{mathtools}
\usepackage[authoryear]{natbib}
\begin{document}

\begin{titlepage}


\title{Modeling community structure and topics in dynamic text networks}
 \author{Teague R. Henry\\
  \textit{University of North Carolina at Chapel Hill} \\ David Banks \\
   \textit{Duke University}  \\Derek Owens-Oas \\\textit{Duke University} \\
   Christine Chai \\\textit{Duke University}}
 \date{Submitted to \textit{Journal of Classification}}




\maketitle
\begin{abstract}
The last decade has seen great progress in both dynamic network
modeling and topic modeling.
This paper draws upon both areas to create bespoke Bayesian model applied to a dataset consisting of the top 467 US political blogs in 2012, their posts over the year, and their links to one another. Our model allows dynamic topic
discovery to inform the latent network model and the network structure
to facilitate topic identification.
Our results find complex community structure within this set of blogs, where
community membership depends strongly upon the set of topics in which
the blogger is interested. We examine the time varying nature of the Sensational Crime topic, as well as the network properties of the Election News topic, as notable and easily interpretable empirical examples.
\end{abstract}



\end{titlepage}

\section{Introduction}
\label{S:1}

Dynamic text networks have been widely studied in recent years,
primarily because the Internet stores textual data in
a way that allows links between different documents. 
Articles on the Wikipedia \citep{hoffman2010online}, citation networks 
in journal articles \citep{moody}, and linked blog posts \citep{latouche} 
are examples of dynamic text networks, or networks of documents that are generated over time. 
But each application has idiosyncratic features, such as the structure of the links and the nature of the time varying documents, so analysis typically requires
bespoke models that directly address those aspects.

This article studies dynamic topic structure and the network properties of the top 467 US political
blogs in 2012.
Some key features of this data set are (1) topics, such as presidential election news, that evolve over time and
(2) community structure among bloggers with similar interests.
We develop a bespoke Bayesian model for the dynamic interaction between text
and network structure, and examine the dynamics of both the discourse 
and the community structure among the bloggers.

Our approach combines a \textit{topic model} and a \textit{network model}. 
A topic model infers the unobserved topic assignments of a set 
of documents (in this case, blog posts) from the text. 
And a network model infers communities among the nodes (in this case, 
blogs that tend to link to one another).
In combination, we find blocks of blogs that tend to post on the same 
topics and which link with one another. 
These blocks, which we call \textit{topic interest blocks}, allow 
one to examine sets of similar blogs, such as those that post only on the 
2012 election or those that are only interested in both the Middle East and 
foreign policy.
Topic interest blocks allow text content to guide community discovery 
and link patterns to guide topic learning.

We begin with a review of terminology in topic modeling.
A \textit{corpus} is a collection of \textit{documents}. 
A document, in our case a post, is a collection of \textit{tokens}, 
which consist of words and \textit{n-grams}, which are sets of words 
that commonly appear together (``President of the United States'' is a common
5-gram). 
In our application, a \textit{blog} produces posts. 
A \textit{topic} is a distribution over the tokens in the corpus.
Typically, a post concerns a single topic.
One such topic might be described as ``the 2012 election'', but this
labeling is usually done subjectively, after estimation of the topic
distributions, based on the high-probability tokens.
For example, the 2012 election topic might put high probability
on ``Gingrich'', ``Santorum'', ``Cain'' and ``primaries''\footnote{Gingrich, Santorum and Cain all refer to candidates in the 2012 Republican presidential primary.}.

An early and influential topic model is Latent Dirichlet Allocation
(LDA), proposed in \citet{Blei2003}. 
It is a bag-of-words model, since the order of the tokens is ignored.
LDA assumes that the tokens in a document are drawn at random from
a topic.
If a document is about more than one topic, then the tokens are
drawn from multiple topics with topic proportions that must be estimated.
The LDA generative model can produce a document that is 70\% about the
2012 election topic and 30\% about a Supreme Court topic by repeatedly 
tossing a coin with probability 0.7
of coming up heads.  
When it is heads, LDA draws a word from the 2012 election distribution;
otherwise, it draws from the Supreme Court distribution.
Markov chain Monte Carlo allows one to reverse the generative model, 
so that given a corpus of documents, one can estimate the distribution
corresponding to each topic, and, for each document, the proportion of that
document that is drawn from each topic.

In our application, topic specific word probabilities evolve over time---the token
``Gingrich'' is more probable early in 2012 than later, when he dropped out. 
\citet{blei2006dynamic} develops a method that allows for topic drift, 
so the probability of a token in a topic can change (slowly) through
an auto-regressive process.
But blog data requires the possibility of rapid change; ``Benghazi''
did not occur in the corpus before September 11, but thereafter
was a high-probability token. 
We develop a dynamic version of a topic model described in
\citet{yin}. The way we infer topics allows for both slow drift and the sudden appearance 
of new words, and even new topics, over the course of the year. 

There is a second source of information in the blog data that previous dynamic 
topic models cannot utilize. 
It is the links between blogs, which prompt a network model.
Here a blog is a \textit{node}, and a hyperlink between blogs
is an \textit{edge}.
We use an exponential random graph model \citep{holland,Wasserman1996a} 
to estimate the probability of an edge through a logistic regression 
on predictors that include node characteristics and other explanatory 
variables. 
This framework can be combined with clustering methods to
perform \textit{community detection}, where a community is a set of
nodes that are more likely to create edges among themselves than with
other nodes.

There are a number of recent methods for community detection. 
One is a family of algorithms that use modularity optimization
\citep{Newman2004a}. 
But the corresponding models are not parametric and do not support
Bayesian inference. 
A popular alternative is the latent space model of \citet{Hoff2002}. 
It estimates nodel locations in an unobserved space which then defines 
the community structure; but it is too computationally demanding for the large blog posts data set. 

We prefer the \textit{stochastic block model} of \citet{snijders97}. 
Stochastic block models place nodes into latent communities based
on the observed pattern of links between nodes, which are modeled 
using independent Bernoulli random variables. 
It has been extended as the mixed membership block model
\citep{Airoldi2008b}, which allows nodes to be members of more than 
one community.  
In that spirit, the model developed in this paper keeps the stochastic 
block modeling framework, but permits nodes to have idiosyncrasies 
in their connection patterns that are not solely due to community 
membership, but also reflect node covariates (in this application, 
the degree of the blogs' interests in specific topics). 
Shared community membership increases edge formation probability, and nodes 
in different communities that have shared topic interests also have
elevated probabilities of linking. A stochastic block model can be easily expressed within an exponential random graph modeling framework

Combining topic information and linkage information through the 
topic interest blocks is our key methodological contribution in this article. A secondary contribution is extending the topic model of \citet{yin} into a dynamic topic model.
Researchers have started to develop models that combine network
analysis and topic analysis, mostly in the context of static networks.
\citet{chang} describes a relational topic model in which the
probability of links between documents depends upon their topics
and applies it to two datasets of abstracts and a set of webpages
from computer science departments.
\citet{ho} applies such methods to linked hypertext and citation networks.  
\citet{WangD} develops a model for the case in which there are
noisy links between nodes, in the sense that there are links between
documents whose topics are not related. 
\citet{yin} does related work on clustering documents through use
of a topic model.  However none of these methods allow for the
simultaneous modeling of dynamic topics with a community structure 
on the nodes.

Our model uses text and covariate information on each node to 
group blogs into blocks more likely to post on the same topics and link to one another.
This approach expands upon community detection, but also fundamentally alters how communities are defined. 
We assume that if two blogs are interested in the same topics, then they
are more likely to link to each other and form a community. 
Estimating the extent to which blogs post about the same topics helps explain community structure, above and 
beyond the community structure described by linkage pattern.  
Furthermore, integrating community detection into topic models  
allows the linkages to inform the allocation of topics, connecting 
network structure to topic structure. 
So inference on topic distributions is supplemented by non-text information. This results in communities that are defined both on the pattern of links (traditional community detection), as well as textual data.
One consequence of this approach is that communities are more grounded in the substantive reason for any community structure, shared interest in various topics. 

In particular, for the 2012 blog application, we wanted a bespoke Bayesian 
model that (1) allows topic distributions to change over time, both
slowly and quickly, (2) classifies blogs into blocks that share topic
interests and have elevated internal linkage probabilities, and 
(3) theoretically enable use of covariate information on blogs. This includes prestige, sociability, whether recently linked, and more. Here we extend covariates from static stochastic block models, as 
in \citet{faust}, to dynamic networks.
Some covariates are fixed (e.g., topic interests) whereas others are time-varying 
(e.g., whether recently linked).

Section 2 describes our dataset and its preparation.
Section 3 gives a generative dynamic model for both the text and the network.  
Section 4 specifies the Bayesian prior and posterior inference algorithm used to estimate model parameters. 
Finally, in Section 5, we present several findings from the political 
blog data, and Section 6 finishes with a discussion of possible 
generalizations.

\section{Political Blogs of 2012}

Our data consists of the blog posts from the top 467 US political blogs for the year 2012, as
ranked by \citet{tech}. 
This dataset has a dynamic network structure since blog posts
often link to each other, responding to each other's content.
Additionally, the topic structure of the blog posts reflect 
different interests, such as the presidential campaign or 
sensational crime. 
The token usage in each topic changes over time, sometimes quite suddenly, 
as with the appearance of the tokens ``Trayvon'' and ``Zimmerman''\footnote{George Zimmerman shot and killed Trayvon Martin in March of 2012.} 
in March, 2012, and sometimes more gradually, as with the slow fade
of the token ``Gingrich''\footnote{Newt Gingrich gradually faded to political irrelevance after a failed presidential primary run.} during the spring.  
Over the 366 days in 2012, a leap year, the political blogs accumulated
109,055 posts.

\subsection{Data Preparation}

Our data were obtained through a collaboration with MaxPoint Interactive, 
now Valassis Digital, a company headquartered in the Research Triangle that 
specializes in computational advertising. 
Using the list of 467 U.S.\ political blog sites curated
by \texttt{Technorati}, computer scientists at MaxPoint scraped all the text
and links at those sites (after declaring robot status and following
all robot protocols). 

The scraped text was stemmed, using a modified version of Snowball
\citep{mcnamee} developed in-house at MaxPoint Interactive. 
The initial application removed all three-letter words, which
was undesirable, since such acronyms as DOT, EPA and NSA are
important. That problem was fixed and the data were restemmed. 

The second step was filtering. 
This filtering was based on the variance of the unweighted term-frequency, 
inverse document frequency (TF-IDF) scores \citep{ramos}. 
The TF-IDF score for token $w$ in blog post $d$ is 
\begin{equation}
\mbox{TF-IDF}_{wd} = f_{wd}/n_w
\end{equation}
where $f_{wd}$ is the number of times that token $w$ occurs in
blog post $d$, and $n_w$ is the number of posts in the corpus
that use token $w$.
Words that have low variance TF-IDF scores are such words as
``therefore'' and ``because,'' which are common in all posts. 
High-variance scores are informative words that are used often in a small
number of posts, but rarely in other posts, such as 
``homosexual'' or ``Zimmerman''. 
Interestingly,
``Obama'' is a low-variance TF-IDF token, since it arises in nearly all
political blog posts.  

Next, we removed tokens that were mentioned in less than 0.02\% of the posts. 
This reduced the number of unique tokens that appeared in the corpus,
as these were unlikely to be helpful in determining the topic token 
distribution across all posts.  
Many of these were misspellings; e.g., ``Merkle'' for ``Merkel'', the 
Chancellor of Germany. Overall, these misspellings were either rare (as in the case of ``Merkle-Merkel''), or incomprehensible. 

After all tokens were filtered, we computed the $n$-grams, starting
with bigrams.
A bigram is a pair of words that appear together more often than
chance, and thus correspond to a meaningful phrase.
For example, the words ``white'' and ``house'' appear in the blog
corpus often, in many different contexts (e.g., race relations and 
the House of Representatives).
But the phrase ``White House'' refers to the official residence of
the president, and appears more often than one would
predict under an independence model for which the expected number
of phrase occurrences is $N p_{\mbox{white}} p_{\mbox{house}}$, where
$N$ is the total amount of text in the corpus and $p_{\mbox{white}}$
and $p_{\mbox{house}}$ are the proportions of the text that are
stemmed to ``white'' and ``house''. 
Bigrams were rejected if their significance probability was greater
than 0.05. 
In examining the bigram set generated from this procedure, it appeared 
to be too liberal; English usage includes many phrases, and about 70\% 
of tested bigrams were retained. 
Therefore we excluded all bigrams occurring less than 500 times corpus-wide. 
This significantly reduced the set of bigrams. 

After the bigrams were computed and the text reformatted to
combine them, the bigramming procedure was repeated. 
This produced a set of candidate trigrams (consisting of a 
previously identified bigram and a unigram), as well as a set 
of candidate quadrigrams (made up of two previously accepted bigrams). 
These candidates were retained only
if they had a frequency greater than 100. This cut-off removed the
majority of the candidate trigrams and quadrigrams. The final vocabulary
consisted of 7987 tokens.  

It is possible to go further, finding longer $n$-grams, but we did 
not.
However, we identified and removed some long $n$-gram pathologies, such 
as the one created by a blogger who finished every post by quoting
the Second Amendment.
There is a large literature on various $n$-gramming strategies \citep{brown}.
Our work did not employ sophisticated methods, such as those that
use information about parts of speech. 
After this preprocessing complete, we had the following kinds of 
information:
\begin{itemize}
\item Stemmed, tokenized, reduced text for each post, the date on which
the post was published, the blog the post it was published on, and links 
to other blogs in the network. 
\item Blog information, including the web domain, an estimate of its prestige
from \texttt{Technorati}, and sometimes information on political affiliation. 
\end{itemize}

From this information, we want to estimate the following:
\begin{itemize}
\item Time evolving distributions over the tokens, where the time evolution 
on a token may be abrupt or gradual.
\item The topic of each post---our model assumes that a post is about a single
topic, which is usually but not always the case (based upon preliminary work
with a more complicated model). 
\item The topic interest blocks, which are sets of blogs that tend to link
among themselves and which tend to discuss the same topic(s).
\item The specific topics of interest to each of the topic interest blocks.
\item The linking probabilities for each pair of blogs, as a function of 
topic interest block membership and other covariates.
\item Posting rates, as a function of blog covariates and external news
events that drive discussion.
\end{itemize}
We now describe the generative model that connects dynamic topic models
with network models in a way that accounts for the unique features of 
this data set.

\section{Model}

The generative model can be described in two main phases: initialization of 
static quantities, such as blogs' topic interest block membership, 
and generation of dynamic quantities, specifically posts and links.
First, the model creates $k$ topic distributions that are allowed 
to change over time. 
Next, it generates time-stamped news events for each topic. 
Each blog is randomly assigned to a topic interest block. 
With these elements in place, post and link generation proceeds. 
For each blog, on each day the number of posts from each topic is generated, 
in accordance to the topic interest block of that blog. 
The content of the post is generated from the day-specific topic 
distribution, and links are generated so as to take account the blog's 
topic interest block.
We now describe each step in the generative model in more detail.

\subsection{Topic and Token Generation}
We begin with the topic distributions, which must
allow dynamic change.
For the $k$th topic, on a specified day $t$, we assume the token 
probabilities ${\bf{V}}_{kt}$ are drawn from a Dirichlet distribution prior. 
This set of topic-specific token probabilities is the 
\textit{topic distribution} on day $t$.
To encourage continuity across days, we calculate the average of topic 
$k$'s topic distribution ${\bf{V}}_{k(t-1):(t-\ell)}$ from the previous 
$\ell$ days and use it as the concentration parameter for the Dirichlet 
distribution from which the present day's topic ${\bf{V}}_{kt}$ is drawn. 
The sampling proceeds in sequence, first calculating each topic's concentration parameter as in \ref{conc} and then sampling each topic as in \ref{dteq} and then moving to the next day. This procedure repeats for times $t = 1:T$. The topics are then distributed:

\begin{equation}
\label{conc}
{\boldsymbol{a}}_{kt} = \frac{1}{\ell} \sum_{t'=1}^{\ell} {\bf{V}}_{k(t-t')},
\end{equation}

\begin{equation}
\label{dteq}
{\bf{V}}_{kt} \sim {Dir}_{|W|}({\bf{a}}_{kt}).
\end{equation}

\subsubsection{Topic Event Generation}

To capture the event-driven aspect of blog posting, we generate events 
which then boost the post rate on the corresponding topic. 
For each topic $k$, at each time $t$, there is some probability $\eta_k$ of 
an event occurring. 
One can choose $\eta_k = .01$ for all $k$, which suggests each topic has 
on average $1$ event every $100$ days. 
Alternatively, different topics can be given different daily event 
probabilities or one can put a prior on $\eta_k$. 
Given $\eta_k$, the daily, topic-specific event indicators are sampled as:

\begin{equation}
E_{kt} \sim Bern(\eta_k).
\end{equation}

When an event happens on topic $k$, blogs with interest in topic $k$ have their posting rates increase by a factor determined by $\psi_k$. Speculating that some topics have events which are much more influential than others, we let this multiplier be topic specific:

\begin{equation}
\psi_{k} \sim Gam(a_{\psi}, b_{\psi}).
\end{equation}

\subsubsection{Block and blog specific topic interest specification}
\added[id=TRH]{With our topic distributions and topic specific events generated, we can now assign blogs to topic interest blocks.} We begin by defining the block-specific topic-interests matrix ${\bf{I}}$, where each column $b$ indicates which of the $k$ topics are of interest  to block $b$. The first ${K \choose 1}$ columns correspond to the singleton blocks, which are interested only in topic $1$, topic $2$, up through topic $K$, respectively. The next ${K \choose 2}$ columns define doublet blocks, which have interest in all of the possible topic pairs. The next ${K \choose 3}$ columns correspond to blocks which have interest in exactly 3 topics, and the final column is for the block which has interest in all $K$ topics:

\setcounter{MaxMatrixCols}{20}
\begin{equation} 
\label{topicblock}
I_{kb} =
\left[ 
\begin{array}{ccccc:ccccc:ccccc:c}
1 & 0 & 0 & \dots & 0 & 1 & 1 & 1 & \dots & 0 & 1 & 1 & 1 & \dots & 0 & 1 \\
0 & 1 & 0 & \dots & 0 & 1 & 0 & 0 & \dots & 0 & 1 & 1 & 1 & \dots & 0 & 1 \\
0 & 0 & 1 & \dots & 0 & 0 & 1 & 0 & \dots & 0 & 1 & 0 & 0 & \dots & 0 & 1 \\
0 & 0 & 0 & \dots & 0 & 0 & 0 & 1 & \dots & 0 & 0 & 1 & 0 & \dots & 0 & 1 \\
0 & 0 & 0 & \dots & 0 & 0 & 0 & 0 & \dots & 0 & 0 & 0 & 1 & \dots & 0 & 1 \\
0 & 0 & 0 & \dots & 0 & 0 & 0 & 0 & \dots & 0 & 0 & 0 & 0 & \dots & 0 & 1 \\
\vdots & \vdots & \vdots & \dots & \vdots & \vdots & \vdots & \vdots & \dots & \vdots & \vdots & \vdots & \vdots & \dots & \vdots & \vdots \\
0 & 0 & 0 & \dots & 0 & 0 & 0 & 0 & \dots & 0 & 0 & 0 & 0 & \dots & 0 & 1 \\
0 & 0 & 0 & \dots & 0 & 0 & 0 & 0 & \dots & 0 & 0 & 0 & 0 & \dots & 1 & 1 \\
0 & 0 & 0 & \dots & 0 & 0 & 0 & 0 & \dots & 1 & 0 & 0 & 0 & \dots & 1 & 1 \\
0 & 0 & 0 & \dots & 1 & 0 & 0 & 0 & \dots & 1 & 0 & 0 & 0 & \dots & 1 & 1 
\end{array}
\right]
\end{equation}

To assign blogs to blocks, we sample their membership with a single draw from a multinomial distribution. This means each blog is a member of only a single block, characterized by the topic interests in the above matrix. Each block assignment is then drawn from a multinomial distribution:

\begin{equation}
b_i \sim Mult(1,{\bf{p}}_B).
\end{equation}

One can choose the probabilities of belonging to each block uniformly, by setting each element of $p_b = 1/B$ where $B = {K \choose 1} + {K \choose 2} + {K \choose 3} + 1$ gives the total number of blocks. Another approach is to partition the probabilities vector into the singlets, doublets, triplets, and all-topics blocks, and allocate probability uniformly to each of these categories, and then uniformly divide up the probability among blocks within each category:

\begin{equation}
{\bf{p}}_B = \begin{pmatrix} {\bf{p}}_1 \\ {\bf{p}}_2 \\ {\bf{p}}_3 \\ p_K \end{pmatrix}, \text{ with }
{\bf{p}}_1 = \begin{pmatrix} p_{1,1}\\ p_{1,2} \\ \vdots \\ p_{1,{\binom{K}{1}}}\end{pmatrix},
{\bf{p}}_2 = \begin{pmatrix} p_{2,1}\\ p_{2,2} \\ \vdots \\ p_{2,{\binom{K}{2}}}\end{pmatrix}, 
{\bf{p}}_3 = \begin{pmatrix} p_{3,1}\\ p_{3,2} \\ \vdots \\ p_{3,{\binom{K}{3}}}\end{pmatrix}, 
p_K = p_{K,{\binom{K}{K}}}.
\end{equation}

For notational convenience throughout the rest of the paper, we define ${\bf{B}}_i$ to be the set of topics which are of interest to blog $i$:

\begin{equation}
\label{blockset}
{\bf{B}}_i = \{k: I_{k b_i} = 1\}.
\end{equation}

With each blog's topic interest indicators known, we can generate blog-specific topic-interest proportions. For example, two blogs may be in the block with interest in topic 1 and topic 2, but one may have interest proportions $(.9, .1)$ while the other has $(.5,.5)$. As is conventional in topic modeling, topic (interest) proportions are drawn from a Dirichlet distribution, though we make the distinction that each blog has a specific set of hyperparameters ${\boldsymbol{\alpha}}_i$. An individual topic interest vector $\boldsymbol{\pi}_i$ is then a draw from a Dirichlet distribution: 

\begin{equation}
{\boldsymbol{\pi}}_i \sim {Dir}_K({\boldsymbol{\alpha}}_i).
\end{equation}

The hyperparameters are chosen such that a blog with interest in topics 1 and 2 is likely to have most of its interest in those topics, though it allows for interest in other topics to occur with small probabilities:

\begin{equation}
{\boldsymbol{\alpha}}_i = \begin{pmatrix} \alpha_{i1} \\ \alpha_{i2} \\ \vdots \\ \alpha_{iK} \end{pmatrix}, \text{ with }
\alpha_{ik} = P 1(k \in {\bf{B}}_i) + 1(k \notin {\bf{B}}_i).
\label{Block}
\end{equation}

\subsubsection{Post Generation}
Given the blogs' topic interest and block membership, along with the event distribution, we can now generate the number of posts a blog produces on a particular topic. Each blog may post on multiple topics, but each post is associated with a single topic.  Every blog has a baseline posting rate which characterizes how active it generally is on days without events. For blog $i$ the baseline post rate $\rho_i$ is sampled from the following distribution:

\begin{equation}
\rho_i \sim Gam(a_{\rho}, b_{\rho}).
\end{equation}

With the blog specific baseline post rate $\rho_i$, the blog specific topic interest proportions $\pi_{ik}$, the topic specific daily event indicators $E_{kt}$, and topic specific post rate multipliers $\psi_k$ accounted for, we construct the expected post rate for each topic, on each blog, each day:

\begin{equation}\label{pois}
\lambda_{tki}  = \rho_i\pi_{ik} + \rho_iE_{tk}\psi_k.
\end{equation}

Given this post rate, the count $D_{tki}$ of posts about topic $k$, on blog $i$, on day $t$ are generated:

\begin{equation}
\label{poislik}
D_{tki} \sim Pois(\lambda_{tki}).
\end{equation}

In the observed data, we don't know the post counts $D_{tki}$ on each topic, but instead we know the marginal counts $D_{ti}$. These are referenced throughout the inference procedure described in section \ref{estimation} and are calculated: 

\begin{equation}
D_{ti} = \sum_{k=1}^K D_{tki}.
\end{equation}

With daily topic specific post counts and token probabilities available, the posts can be populated with tokens. We first sample a total number of tokens for each post. In particular, on day $t$, the token count $W_{tkid}$ for post $d$ about topic $k$ on blog $i$ is sampled:

\begin{equation}
W_{tkid} \sim Pois(\lambda_D).
\end{equation}

Where $\lambda_D$ is the average number of tokens over all posts. The $W_{tkid}$ tokens can then be sampled from the appropriate day and topic specific multinomial distribution with probability vector ${\bf{V}}_{kt}$. This is done for all of the posts in the corpus like so:

\begin{equation}
N_{tkid}^w \sim Mult(W_{tkid}, {\bf{V}}_{kt}).
\end{equation}

\subsubsection{Network Generation}
Finally, we generate the network of links between blogs. Rather than modeling link generation at a post level, we model it at a daily blog to blog level. Specifically, we model a directed adjacency matrix $A_t$ of links, with entry $a_{ii't}$ indicating whether any posts from blog $i$ have links to blog $i'$ on day $t$. The binary logistic regression is suitable for this scenario. We assume the link probability $p_{ii't} = p(A_{ii't} = 1)$ depends on the following factors:

\begin{itemize}
\item $B(i,i') = 1({b}_i = b_{i'}) + {\boldsymbol{\pi}}_i^T {\boldsymbol{\pi}}_{i'}1(b_i \ne {b}_{i'})$ is the similarity (in topic interests) for nodes $i$ and $i'$, and is in the interval [0,1], taking value 1 if and only if blogs $i$ and $i'$ are in the same block. 
\item $L_{i'it} = 1((\sum_{t' = t-7}^{t-1} a_{i'it'}) > 0)$ indicates if blog $i'$ has linked to blog $i$ within the last week (previous to the current time t).
\item $I_{i't} = \frac{1}{t-1}\sum_{t'=1}^{t-1} \sum_{i} a_{ii't'}$ is the average indegree (through time t-1) of the receiving node i'.
\item $O_{it} = \frac{1}{t-1}\sum_{t'=1}^{t-1} \sum_{i'} a_{ii't'}$ is the average outdegree (through time t-1) of the sending node i.
\end{itemize}

The first covariate is sampled and constructed in equations \ref{topicblock}-\ref{Block}, and the other three covariates are defined and calculated as statistics of the past data $\{A_{t'}\}_{t'=1}^{t-1}$. Together with an intercept, they comprise the regressors in a logistic regression for links, which can be written as in Equation \ref{neteq},
 
\begin{equation}
\label{neteq}
\log (\frac{p_{ii't}}{1-p_{ii't}}) = \theta_0 +
\theta_1 B(i,i') + \theta_2 L_{i'it} + \theta_3I_{i't} + \theta_4 O_{it}.
\end{equation}

We specify a normal prior for the intercept and the regression coefficients:

\begin{equation}
\theta_p \sim Norm(\mu_{\theta}, \sigma_{\theta}^2).
\end{equation}

We can use the logistic function to write the probability of a link as:

\begin{equation}
\label{probeq}
p_{ii't} = p(A_{ii't} = 1) = \frac{\exp({\boldsymbol{\theta}}^T {\bf{S}}_{ii't})}{1+\exp({\boldsymbol{\theta}}^T {\bf{S}}_{ii't})},
\end{equation}

with coefficients and covariates written as:

\begin{equation}
\label{coefcov}
{\boldsymbol{\theta}} = (\theta_0, \theta_1, \theta_2, \theta_3, \theta_4)^T \text{ and } {\bf{S}}_{ii't} = (1, B(i,i'), L_{i'it}, I_{i't}, O_{it})^T.
\end{equation}

This form makes it more clear how the model can be cast within the ERGM framework \citep{holland}. Covariates are time dependent as in TERGM literature \citep{krivitsky}. An important note is that covariates depend only on past linking data, which makes this a predictive model of links. Finally, we sample each link as a single Bernoulli trial with the appropriate probability as defined in Equation \ref{probeq}:

\begin{equation}
A_{ii't} \sim Bern(p_{ii't}).
\end{equation}

This generative model for the links can be thought of as a variant of stochastic block modeling \citep{snijders97}, where block membership is ``fuzzy''. In our model, while members of the same block will have the highest probability of linking with other members of the same block, individuals who share similar topic interests, but do are not in the same block are more likely to link than individuals who share no topic interests. This allows for a pattern of linkages that more accurately reflect the empirical phenomena of topic based blog hyperlinks. 

At the end of data generation we have $\{{\bf{B}}_i\}_{i=1}^I$ and ${\boldsymbol{\pi}}_i$ giving the topic interest set and topic interest proportions, respectively, for blog $i$; $K \times T$ matrix ${\bf{E}}$ with daily topic specific event indicators; $K \times I \times T$ array ${\bf{D}}$ with entry $D_{kit}$ giving the number of posts about topic $k$ on blog $i$ at time $t$; $|W| \times K \times T$ array ${\bf{V}}$ of daily topic specific token probabilities; and multidimensional object ${\bf{N}}$ containing the count $N_{tkid}^w$ for each token $w$ in the $d$th post about topic $k$ on blog $i$ at time $t$.

With a theoretically justified data generating mechanism in place, we proceed to Section $4$ to ``invert the generative model'' and derive posterior inference for the parameters of interest.

\section{Estimation}
\label{estimation}
As our dataset of blog posts consists of posts, time stamps, which blog 
posted each post, and each post's links to other blogs, our inferential 
model needs to estimate a number of quantities. 
This section gives the details of how we estimate quantities of 
interest from the data and how we specified our priors.
The notation is dense, so the following guide is helpful.
\begin{itemize}
\item \textbf{Each post's topic assignment}. We observe the content of 
each post, but do not know the topic assignment of each post. 
This must be inferred. We denote this estimate as $z_d$ for post $d$.
\item \textbf{Topic distributions}. We do not know what the content of 
each topic is, or how each topic changes over time. 
We use $\textbf{V}_t$ for the topic token-distributions matrix, and 
for specific topics, we denote this as $\textbf{V}_{kt}$.
\item \textbf{Events and Post Rate Boosts}. Events are not observed and 
must be inferred. 
This $T \times K$ matrix is $\textbf{E}$. 
The event-caused, topic-specific boosts in post rate $\psi_k$ are also inferred.
\item \textbf{Blog specific parameters}. A blog's average post rate and 
topic interests must be inferred. 
The blog average post rate is denoted $\rho_i$, and the topic 
interest proportions is a vector of length $K$, denoted ${\boldsymbol{\pi}}_i$.
\item \textbf{Blogs' block membership}. A Blog's block membership is 
inferred using the linkage pattern and topic assignments of each of the 
blog's posts. 
The $i$th blog's block membership is denoted as $b_i$ and its 
corresponding topic interests indicator vector is ${\bf{B}}_i$.
\item \textbf{Network parameters}. The network parameters govern the 
probability of linkage.
These are depend upon block membership, lagged reciprocity, 
indegree and outdegree through a logistic regression whose coefficients
must be estimated. 
These five network parameters (including intercept) are denoted 
$\theta_0, \theta_1, \theta_2,\theta_3$ and $\theta_4$, respectively.
\end{itemize}




\subsubsection{Hyper Parameter Specification}
The model requires that several parameters be specified \textit{a priori}. In this subsection we describe these hyper parameters in general terms, while in section 5.1, we show which specific values we used to analyze the political blog data.  
The first hyper parameter is $K$, the total number of topics. 
In principle, one could place an informative prior on the number of topics and 
use the posterior mean determined by the data.
This is, however, computationally cumbersome and so we make the 
decision to specify the number of topics in advance 
This approach is used in \cite{blei2006dynamic} and
\cite{Blei2003} for the Latent Dirichlet Allocation models . One can use penalized likelihood as a selection
criterion, as described in  \cite{yin}, or an entropy based
criterion, such as the one described in \cite{arun}.  
We chose the number of topics by running models with different values
of $K$ and selecting the number of topics using the entropy based
criterion of \cite{arun}. 

The time lag $\ell$ for topic dependency needs to be
specified. 
This time lag determines the scale of the topics, and has units in number of days. This determines how long tokens remain in a topic, and can be conceptualized as a smoother over the time changing vocabulary. Smaller values of $\ell$ will produce more varied topic distributions over time, while larger values will reflect slower shifts in the topic content. 
For the node specific parameters, only $P$, the Dirichlet concentration parameter
on the topics which are of interests to a block, is needed. This parameter governs how often blogs are allowed to post outside of topics they are interested in, with lower values allowing for more out of interest posting, and higher values corresponding to restricted topic interests.
Finally, for any reasonable number of topics, a restriction on the
block structure is required  to ensure computational feasibility.  
For an unrestricted block structure with $K$ topics, the total number
of possible blocks that must be evaluated is $\sum_{i = 1}^K {K
  \choose i}$, which is computationally intractable for moderate $K$.  
In this paper, we restrict blocks to have 1,
2, or 3 topic interests, and allow one block to have interest in all topics. 
Finally, we specify the expected number of non-zero topic interest blocks using the prior $\lambda_B$.

\subsection{A Simple Data Augmentation}
While the generative model assumes Poisson distribution on post counts $D_{kit}$, we rely on a data augmentation for the inference procedure. Because counts $D_{it}$ of posts on each blog each day are already known, we augment the generative model with latent variables $\{z_{d_{it}}\}_{d_{it}=1}^{D_{it}}$ which instead tell the latent topic assignment of post $d_{it}$. We can then re-write the Poisson likelihood $\prod_{k=1}^K Pois(D_{kit} | \lambda_{kit})$ as a multinomial likelihood $\prod_{d_{it} = 1}^{D_{it}} Mult(z_{d_{it}} | 1, {\boldsymbol{\xi}}_{it})$ with $\xi_{kit} = \frac{\lambda_{kit}}{\sum_{k=1}^K \lambda_{kit}}$. This reformulation enables use of the topic assignment inference algorithm from GSDMM.

\subsection{Metropolis Within Gibbs Sampling}
We use a Metropolis within Gibbs sampling algorithm \cite{gilks} 
to obtain posterior 
distributions for the parameters defined in the generative model.
This approach consists of four stages: 
\begin{enumerate}
\item   Each day $t$, for each blog $i$, sample a topic assignment $z_{d_{it}}$ for each post $d_{it}$ and update the matrix
of daily topic specific token-distributions $\textbf{V}_t$.
\item   For blogs, update topic interest proportions ($\pi_{ik}$), and base rate for posting ($\rho_i$). For events, update the event matrix $\textbf{E}$, and activation level parameters $(\psi_k)$.
\item  Update the network parameters, i.e., $\theta_0, \; \theta_1, \; \theta_2, \;
  \theta_3$ and  $\theta_4$. 
\item  Update each blog's block assignment $b_i$ and corresponding topic interest indicators ${\bf{B}}_i$.
\end{enumerate}

\subsection{Topic Modeling and Post Assignment}

Both posts' topic assignments as well as the topic distributions themselves are unobserved and must be inferred. A preferred algorithm for inferring post topic assignment and topic token-distributions would first assign each post a single topic, and second have some flexibility in collapsing topic distributions together. 

To those ends, we adapt the Gibbs Sampler for the
Dirichlet Mixture Model (GSDMM) of \cite{yin}. 
As originally proposed, the GSDMM classifies a set of
documents into specific topics. 
The tokens of a post are assumed to be generated from the topic specific multinomial which that post was assigned, and many tokens may be instantiated in multiple topics (e.g., common words such as ``therefore'').
The assignment of each post to a \emph{single topic} differs from the Latent Dirichlet Allocation model,
which models documents as mixtures across a number of topics.  
GSDMM estimates the probability that a document $d$ is about topic 
$k$, given the current topic vocabulary distribution, as   
\begin{equation}
P(z_d = k | \textbf{V}_k, d) = \frac{m_{k,-d} + \alpha}{|D|-1+K\alpha} 
\frac{\prod_{w \in d}\prod_{j = 1}^{N_d^w}(N_{k, -d}^w + \beta + j -1)}{\prod^{N_d}_{i=1}(n_{k,-d} + |W|\beta +i-1)},
\end{equation}
where $m_{k,-d}$ is the number of posts currently assigned to topic $k$ 
(not including the topic assignment of post $d$),  $N_d^w$ is the 
number of occurrences in post $d$ of token $w$, and $N_{k,-d}^w$ is 
the number of occurrences of token $w$ in topic $k$ (not including the content 
of post $d$). 
The $\alpha$ controls the prior probability that a post is assigned 
to a topic; increasing $\alpha$ implies that all topics grow equally likely.
The $\beta$ relates to the prior probability that a token will have relevance 
to any specific topic; increasing $\beta$ results in fewer topics being found 
by the sampler. Finally, $|D|$ is the number of posts in total, and $|W|$ is the size of the vocabulary.

As originally proposed by \cite{yin}, GSDMM is a static model. 
We modify it by allowing ${\bf{V}}$ to vary over time. 
For readability, we suppress the subscripts and denote the
specific post $d_{it}$ by $d$. 
We define 

\begin{equation}
m^*_{k,t, -d} = (\sum_{t' = t-\ell}^t D_{t'k}) - 1 \text{ with }  
D_{tk} = \sum_{i} D_{tki},
\end{equation} to be the number of posts
assigned to topic $k$  in the interval from $t - \ell$ to $t$, 
not including post $d$ by blog $i$ at time $t$.
Also we let
\begin{equation}
N^{*w}_{k,t,-d} =  \sum_{t' = t- \ell}^t N^w_{k,t'},
\label{eq21}
\end{equation} 
be the number of times that token $w$ occurs in topic $k$ in the
interval from $t - \ell$ to $t$, not including post $d$. 
This defines a sliding window that allows the sampler to use information 
from the recent past to infer the topic to which a post belongs, 
while allowing new tokens to influence the assignment of the post at the 
current time point. 
The probability of assigning post $d$ to topic $k$ is then:
\begin{equation}\label{mGSDMM}
\Pb[z_{d} = k \,|\, {\bf{V}}_{kt}, d] = 
\frac{m^*_{k, t, -d} + \alpha}{|D_{t-\ell:t}|-1+K\alpha} 
\frac{\prod_{w \in d} \prod_{s = 1}^{N_{d}^w} (N^{*w}_{k,t, -d} + \beta
  + s - 1)}{\prod^{N_{d}}_{i=1}(N^*_{k,t,-d} + |W|\beta +i-1)},
  \end{equation}

where $|D_{t-\ell:t}|$ is the number of posts within the lag window. Note that (\ref{mGSDMM}) does not use information about the blog that 
generates the post.
So the final step is to incorporate the tendency of blog $i$ to
post on topic $k$ 
at time $t$, using the Poisson rate parameter in (\ref{pois}). 
Using the normalized point-wise product of conditional probabilities, the final
expression for the probability that post $d$ (i.e., $d_{it}$) 
belongs to topic $k$ is
\begin{equation}\label{final}
\Pb[Z_{d} = k \,|\, {\bf{V}}_{k,t}, d, \lambda_{ikt}] = 
\frac{\Pb[Z_{d} = k \,|\, {\bf{V}}_{k,t}, d] \Pb[Z_{d} = k \,|\,
\lambda_{ikt}]}{\sum_{q = 1}^K \Pb[Z_{d} = q \,|\, {\bf{V}}_{q,t}, d]
\Pb[Z_{d} = q \,|\, \lambda_{iqt}]}. 
\end{equation}
To reduce computation, we approximate $\Pb[Z_{d} = k \,|\, \lambda_{ikt}]$ 
with $\lambda_{ikt}/\sum_{j = 1}^K \lambda_{ijt}$, as clarified in the data augmentation above.

The topic assignment of a post can now be Gibbs sampled using equation
(\ref{final}). 
The sampler assigns the first post to a single topic, updates the
topic-token distributions  
with the content of that post, then continues to the next post, and repeats. 
At each time point, the sampler sweeps through the set of posts
several times so that the topic assignments can stabilize. 
The preferred number of sweeps depends on the complexity of the posts on that day, 
but it need not be large.   
After some exploration, this study used 10 sweeps at each time point
per iteration. 

In summary, after topic assignment has been completed for a given day $t$, all posts within that day will have a single topic assignment. Moving to day $t+1$, the topic assignment for all posts that day will utilize the information from day $t$ through that day's topic-specific, token-distribution estimator specified in Equation (\ref{eq21}). Once the topic assignment estimator reaches the final day $T$, all posts will have assigned topics, and all topics will have a time varying token-distribution. The post specific topic assignments are then used in the next step of estimating blogs' topic interest vectors.

\subsection{Node Specific Parameters and Event Parameters}

Once posts are assigned to topics, the next step is to update the
node specific parameters,  
specifically the blog topic interest vector ${\boldsymbol{\pi}}_i$ and the blog posting rate $\rho_i$.  

The topic interest vector is updated in a Metropolis-Hastings step. As is standard in M-H, we specify a proposal distribution, a likelihood, and a prior. The proposal ${\boldsymbol{\pi}}^*_i$ is a draw from a Dirichlet distribution with ${\boldsymbol{\alpha}}_i = {\boldsymbol{\pi}}_i D_i$, where $D_i$ is the total number 
of posts generated by node $i$. 
The likelihood is 
\begin{equation}\label{lik}
\prod_{t = 1}^T \prod_{k=1}^K \mathbb{P}({D_{kit}} | \lambda_{kit}),
\end{equation}
where $\mathbb{P}({D_{kit}}| \lambda_{kit})$ is the Poisson likelihood from equation (\ref{poislik}) representing the day specific number of posts on blog $i$ assigned to each topic. Note the dependence of $\lambda_{kit}$ on ${\boldsymbol{\pi}}_i$ comes through equation (\ref{pois}).
A hierarchical prior is used, which is Dirichlet($\alpha_{B_i}$),
where the parameters are  
defined by the current block assignment of node $i$, as in equation
(\ref{Block}). This step requires estimates of each blog's block assignment, which will be described later.

The $i$th blog's posting rate $\rho_i$ is also updated using a
Metropolis-Hastings step,  
where the proposal distribution is a Normal truncated at 0, with mean
equal to $\rho_i$ and  
standard deviation equal to $\sigma^2_\rho$. 
The likelihood evaluated is the same as in equation (\ref{lik}).  
The prior is a univariate Normal truncated at 0 and with mean
$\rho$ and variance $\sigma^2_\rho$. 
Truncated normal distributions are used to uncouple the mean and the variance.

Next to update are the event matrix ${\bf{E}}$ and activation boost parameters $\psi_k$. The event matrix is updated with a series of Metropolis steps for each time point and topic.  
The proposal is simply $1$ if $E_{k,t} = 1$ and $0$ if $E_{k,t} = 0$. 
The likelihood is the same as equation (\ref{lik}), except, for each topic $k$ at each time $t$, the product of Poisson densities is over the blogs $i=1:I$. The prior is simply a Bernoulli with parameter $E_\pi$.

Each activation parameter $\psi_k$ is updated with a
Metropolis-Hastings step, where the proposal $\psi_k^*$  
is a truncated normal at 0 with mean $\psi_k$ and standard deviation
$\sigma_{\psi}$.  
Again, the likelihood is similar to (\ref{lik}), in that it is the product indexed over time of the Poisson densities for every blogs' number of posts in topic $k$. Unlike the original likelihood, however, the second product is over blogs $i=1:I$.   
The prior distribution on $\psi_k$ is a normal truncated at 0 with
mean $\psi$ and  
standard deviation $\sigma^*_\psi$.

\subsection{Network Parameters}

The network parameter set consists  of the vector
$\boldsymbol{\theta} = (\theta_0, \ldots, \theta_4,)$ as defined in equation
(\ref{neteq}). 
Each network parameter can be sampled using a Metropolis within Gibbs
step. Specifically, the Bernoulli likelihood portion (equivalently a logistic regression) can be expressed as:  
\begin{equation}
\prod_{i \neq j} \prod_{t \in \{1, \dots, T\}} \frac{\exp((\theta_0 +
\theta_1 B(i,j) + \theta_2 L_{jt} + \theta_3I_{j} + \theta_4 O_{i} ))^{A_{ijt}}}{1+\exp(\theta_0 +
\theta_1 B(i,j) + \theta_2 L_{jt} + \theta_3I_{j} + \theta_4 O_{i})}.
\end{equation}
To update each parameter, one conditions on all other pieces of
information in the model. Proposals are normal with mean set to the
current value of the parameter, and a standard deviation specific to
the parameter, while the priors are normal with a given mean and
standard deviation. Note here that this sampling relies on estimates of the block membership of each blog.

\subsection{Block Assignment}

Previously we described the sampling routine for post-topic assignment, topic token-distributions, node specific parameters, event matrix and boosts, and network parameters. The blog specific block memberships remain to be estimated. Due to the complexity of their dependence on many other components of the model, we describe block estimation last.

Recall that a blog's block describes two things. The first is the set of topics the blog will be most likely to post on. The second is a blog will tend to link more to other blogs that are within the same block as it is. Therefore, block assignment for a given node $i$ is informed by several pieces of data. These are its topic interests, the network parameters, other blog's block memberships, and the observed links. One assumption of our model is that a node's network position and
topic interest are conditionally  
independent given block assignment, which in turn makes the sampling
of a block assignment considerably simpler.  
After simplification, a node's potential block assignment is informed by
the number of nodes already assigned to each block. 
Ultimately, the probability that a node $i$ will be assigned to the $b$th block is
proportional to 
\begin{eqnarray}
\label{blockassign}
\Pb[b_i = b \,|\, \textbf{A}, \boldsymbol{\theta}, \boldsymbol{\pi}_i, \textbf{B}_{-i}] \propto \hspace{2.5in} \nonumber \\
\frac{N_{b,-i} +\alpha_B}{\alpha_B|B| + N -1}P(A \,|\, \boldsymbol{\theta}, B_i = b)
P(\pi_i \,|\, B_i = b) P(|B| \,|\, \lambda_B), 
\end{eqnarray}
where $N_{b, -i}$ is the number of nodes assigned to block $b$, not
including node $i$, $\alpha_B$ is  
related to the prior probability of being assigned to any block
(analogously to $\alpha$ in the topic model),  
$\theta$ is the complete set of network parameters, 
$|B|$ is the number of blocks with non-zero membership while 
node $i$ is being considered for potential assignment to block $b$, $\lambda_B$ is the prior number of blocks expected to exist, and $\textbf{B}_{-i}$ is the set of block assignments with the $i$th blog's block assignment removed. As such, the first term acts as a penalty term on the number of blogs in any given block, the second term is the Bernoulli likelihood (logistic regression) of observed links given other block assignments and network covariates, and the third term is the probability of blog $i$'s topic interest vector given the considered block assignment.
Finally $P(|B| \,|\, \lambda_B)$ is the Poisson probability of $|B|$ given
$\lambda_B$ and acts as a penalty term for the number of blocks with non-zero membership. 

To elaborate briefly, the first term in (\ref{blockassign}) and the final term in (\ref{blockassign}) together act as tunable priors on the distribution of sizes of blocks, as well as the number of non-empty blocks. In the generative model, we specify individual priors on the probability of membership in each block. However, for any reasonable $K$, the total number of blocks $B = 1 + \sum_{j=1}^3 {K \choose j}$ exceeds the number
of blogs to assign to blocks. So the block assignment sampler accounts for blocks that have no members. This is a desirable feature since the analyst need not specify exactly how many blocks are instantiated in the model. As compared to the number of possible blocks, the number of non-empty blocks is a rare event. This enables use of the Poisson distribution as a reasonable (and computationally feasible) approximation for the number of non-empty blocks. 

Due to the computational intensity of computing the block specific probabilities, we restrict the number of topics in which a block can have interest. In our work, blocks may be interested in at most three topics, except
for one block that is interested in all topics (to account for such 
blogs as {\em The Huffington Post} or {\em The New York Times}'s
blog).
Furthermore, during sampling, we restrict the blocks considered for a
given node $i$ by only considering blocks that have topics for 
which node $i$ generated at least one post. 
These restrictions change the normalizing constant, though the
relative probabilities of the blocks  
considered remains the same.

\subsection{Summary of Estimation}

The primary goals of the estimation routine are to obtain post-topic assignments, and then topic token-distributions, as well as blog- specific block assignments. These are the main parameters of interest in our application, as they describe the dynamic nature of each content topic and the range of interests and communications each blog has respectively. Alongside this information, we also estimate several other parameters such as those which govern linkage formation, which topics are active when (via events), topic post-rate boost parameters, and blogs' topic interests. While each of these are informative in their own right, in our application below, we chose to focus on the topic distributions and the block assignments of each blog.

\section{Results}

\subsection{Prior Choices}

Changes in the topics for this dataset are expected to be
slow, aside from sudden events that abruptly add new tokens 
(e.g., ``Benghazi'' or ``Sandy Hook'').
Therefore, we used a lag $\ell$ parameter of 62 days to capture gradual 
drift in the topics over time.
Specifically, the distribution over tokens for each topic
was estimated based upon a sliding window for the preceding 62
days.
Within that window, all posts had equal weight. 

\begin{figure}[H]
\includegraphics[width=.7\textwidth]{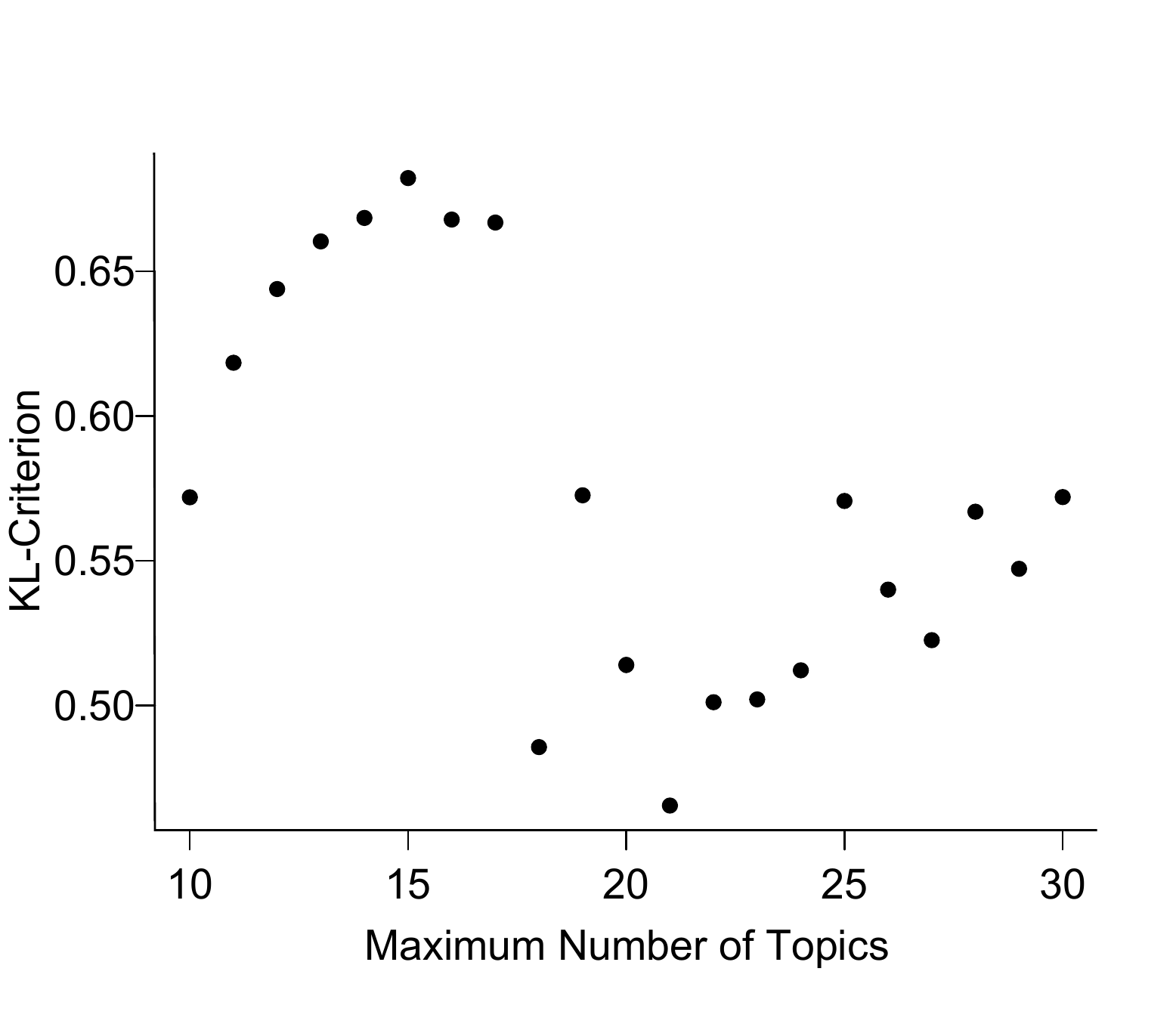}
\centering
\caption{The criterion curve, as in \cite{arun}, for determining
the number of topics.}
\label{KL}
\end{figure}

To determine the number of topics, we used the criterion 
developed by \cite{arun}. The goal of this criterion is to determine the number of topics that lead to the smallest value of the criteria. It is important to note that this criteria is only based off of the topic distribution over the posts, and does not take into account the network structure. This is, of course, a limitation of this criteria and a suggestion for further research. 
Figure \ref{KL} shows the criterion curve, which considers fitting
anywhere between 1 and 30 topics.
The curve has its minimum at 22, and thus our study fixes the
number of topics to be 22.

Once the number of topics is established, the restrictions on the blocks 
and the parameter $P$, as introduced in equation (1), can be set.
Recall that each block may only be interested in 1, 2, 3 or all
topics. 
Finally, $P$, the out of block interest parameter which governs the blogs ability to post on out of interest topics, was set to 50, to allow some freedom for blogs to post 
on topics outside of their block's interests, but nonetheless mostly focus 
on the block's interests. 

The network model specified an edge parameter, a mean
in-degree and out-degree parameter, a 7-day lag parameter, and a block
membership parameter. 
The edge parameter acts as the intercept for the network model. 
Mean in-degree and out-degree are nodal covariates
consisting of the average daily in-degree and out-degree for each
node. 
This allows modeling of differentially popular blogs. 
Finally, to add in temporal dependency, the 7-day lag is an
indicator function that takes the value 1 if and only if the
pair of blogs has been linked within the previous 7 days,
and is otherwise 0 (this captures the fact that bloggers sometimes
have debates, which produce a series of links over a relatively 
short period of time).
Vague priors were set for each of the
network model parameters; all were normals with mean 0 and 
standard deviation 1000. 
The proposal standard deviation was set to 1 for the edge
parameter, and to 0.25 for each of the other parameters in the network
model. 

For the topic model the $\alpha$ and $\beta$ parameters were both set
to 0.1. 
The prior for the average post rates $\rho_i$ in equation (2) 
was a truncated normal at 0, with mean 4 and standard deviation 1000.
The prior for topic activation parameters $\psi_k$ in equation (2) was 
set as a truncated normal at 0 with mean 0 and standard deviation 1000,
and a proposal standard deviation of 0.5. 

Additionally, 25 was set as
the prior mean number of blocks ($\lambda_B$), and the prior tendency
for block membership $\alpha_B$ was set to $1$. The prior probability
of topic activation was set to 0.2. 

The sampler ran for 1000 iterations. To ensure mixing for
the network parameters, at each iteration the network parameters
were updated 10 times. 
During each iteration, there were 10 sub-iterations for the topic 
model and 10 sub-iterations for the block assignments. 
The first 100 overall iterations were discarded as burn-in, 
the remaining 900 were thinned at intervals of 10.   

\subsection{Findings}

The sampler converged to stationarity quickly in every parameter. 
To assess the mixing of the post to topic assignment, and of 
the blogger to block assignment, we calculated Adjusted Rand Indices 
\citep{hubert,steinley} for each iteration $i$ compared to iteration $i-1$. 
The post to topic assignment was very stable, with a mean Adjusted Rand Index of 0.806 and standard deviation 0.047. 
The block assignment was less stable, with a mean Adjusted Rand Index 
of 0.471 and standard deviation 0.031. 
We believe this variability is due to the fact that many
bloggers tended to post on whatever news event captured their
attention, making it difficult to assign them to a block with
interest in no more than three topics.
However, their interests were not so wide that they were reasonably
assigned to the block that is interested in all topics.

All domain level parameters converged successfully. The domain
rate parameter $\rho_i$ was estimated for each domain, and the
posterior means of the domain rates had a mean of 0.632 and a standard
deviation of 1.67. The largest domain rate was 22.69. The distribution
of domain post rates was highly skewed, with few blogs having a very
high average post rate, and most blogs having a lower post rate.  

The topic specific activation parameters $\psi_k$ converged
successfully. Information on the posterior means and standard
deviations is in Table \ref{ExciteParams}, and were calculated after 
the topics had been defined. The topics Election and Republican Primary
have the greatest posterior means, which suggests that these topics
were more event driven than other topics.  

\subsubsection{Topic Results}

The topic model found 22 topics, each of which had distinct 
subject matter. 
Table \ref{specwords} contains the topic titles and total number 
of posts in each topic, as well as the three tokens that have 
the highest predictive probability for that
topic over all days. 
Predictive probability was calculated using Bayes' rule: 
\begin{equation}
P(Z_d = k \,|\, w \in d) =\frac{P(w \in d \,|\, Z_d = k)P(Z_d = k)}{P(w \in d)}.
\end{equation}

Table \ref{popwords} contains the five most frequent tokens in each
topic over all days. 
Topics were named by the authors on the basis of the most predictive 
tokens as well as the most frequent tokens over all days.  
Some of these tokens may seem obscure, but in fact they are generally
quite pertinent to the identified topics.

\begin{landscape}
\begin{table}[]
\centering
\caption{Topic names and their most specific tokens.}
\label{specwords}
\begin{tabular}{llllll}
\hline
\multicolumn{1}{c}{Topic Name} & \multicolumn{1}{c}{\# of posts}  & \multicolumn{3}{c}{Highest Specificity Tokens}                         \\ \hline
\multicolumn{1}{c}{}           & \multicolumn{1}{c}{}            & \multicolumn{1}{c}{1} & \multicolumn{1}{c}{2}  & \multicolumn{1}{c}{3} \\ \hline
Feminism          &  3971                               & russel.saunder.juli   & circumcis              & femin                 \\
Keystone Pipeline              &  4422                               & loan.guarante.program & product.tax.credit     & tar.sand.pipelin      \\
Birth Control                  &  2703                               & contracept.coverag    & birth.control.coverag  & religi.organ          \\
Election                       &  14713                               & soptic                & cheroke                & eastwood              \\
Mortgages                      &  2130                               & estat                 & probat                 & fiduciari             \\
Entertainment                 &  10555                               & email.read.add        & olivia                 & free.van              \\
Middle East             &  6068                               & mursi                 & morsi                  & fatah                 \\
LGBT Rights                    &  5425                               & anti.gay.right        & support.equal.marriag  & equal.marriag         \\
Sensational Crime              &  6423                               & zimmerman             & lanza                  & mass.shoot            \\
Technology                     &  3230                               & mail.feel.free        & pipa                   & ret                   \\
Supreme Court                  &  1767                               & commerc.claus         & bork                   & chief.justic.robert   \\
Bank Regulation                &  5222                               & volcker               & dimon                  & libor                 \\
National Defense      &  1977                               & iaea                  & iranian.nuclear.weapon & warhead               \\
Republican Primary             &  9351                               & poll.mitt.romney      & nation.popular.vote    & romney.lead           \\
Voting Laws                    &  7865                               & ohio.secretari.state  & voter.registr.form     & hust                  \\
Political Theory               &  1448                               & bylaw                 & rawl                   & sweatshop             \\
Eurozone         &  1832                               & standalon             & troika                 & ecb                   \\
Taxation                       &  8435                               & tax.polici.center     & health.care.spend      & top.tax.rate          \\
Diet and Nutrition                         &  3057                               & spielberg             & harlan                 & calori                \\
Education                      &  2909                               & chicago.teacher.union & chicago.public.school  & charter.school        \\
Global Warming                 &  2205                               & arctic.sea.ice        & sea.ice                & sea.level.rise        \\
Terrorism                      &  3347                               & kimberlin             & broadwel               & assang                \\ \hline
\end{tabular}
\end{table}
\end{landscape}

\begin{landscape}
\begin{table}[]
\centering
\caption{The most frequent words in each topic.}
\label{popwords}
\begin{tabular}{llllll}
\hline
Topic Names & \multicolumn{5}{c}{Most Frequent Words} \\ \hline
\multicolumn{1}{c}{} & \multicolumn{1}{c}{1} & \multicolumn{1}{c}{2} & \multicolumn{1}{c}{3} & \multicolumn{1}{c}{4} & \multicolumn{1}{c}{5} \\ \hline
Feminism & women & peopl & dont & person & life \\
Keystone Pipeline & energi & oil & price & compani & industri \\
Birth Control & right & state & law & marriag & women \\
Election & obama & romney & peopl & presid & polit \\
Mortgages & case & court & bank & judg & attorney \\
Entertainment & peopl & dont & good & work & game \\
Middle East & israel & islam & american & peopl & countri \\
LGBT Rights & gay & peopl & marriag & homosexu & support \\
Sensational Crime & gun & polic & report & peopl & zimmerman \\
Technology & compani & googl & facebook & appl & user \\
Supreme Court & law & court & state & case & constitut \\
Bank Regulation & bank & market & money & price & compani \\
National Defense & iran & militari & israel & nuclear & obama \\
Republican Primary & romney & republican & obama & poll & vote \\
Voting Laws & state & vote & elect & voter & counti \\
Political Theory & libertarian & peopl & right & state & govern \\
Eurozone & bank & debt & economi & rate & govern \\
Taxation & tax & state & govern & cut & obama \\
Diet and Nutrition & peopl & dont & govern & polit & work \\
Education & school & student & teacher & educ & state \\
Global Warming & climat & climat.chang & temperatur & scienc & scientist \\
Terrorism & report & govern & attack & inform & case\\
\hline
\end{tabular}
\end{table}
\end{landscape}

\begin{table}[H]
\centering
\caption{Topic Specific Activation Parameters $\psi_k$}
\label{ExciteParams}
\begin{tabular}{lll}
\hline
Topic                     & Posterior Mean & Standard Deviation \\ \hline
Feminism                  & 0.0034         & 0.0029             \\
Keystone Pipeline         & 0.0037         & 0.0017             \\
Birth Control             & .00001         & .00003             \\
Election                  & 0.3917         & 0.0249             \\
Mortgages                 & .00001         & .00005             \\
Entertainment             & 0.0018         & 0.0014             \\
Middle East               & 0.0378         & 0.0129             \\
LGBT Rights               & 0.0028         & 0.0026             \\
Sensational Crime         & 0.0208         & 0.0085             \\
Technology                & 0.0012         & 0.001              \\
Supreme Court             & 0.0022         & 0.0032             \\
Bank Regulation           & 0.0021         & 0.0021             \\
National Defense          & 0.0014         & 0.0013             \\
Republican Primary        & 0.2267         & 0.0188             \\
Voting Laws               & 0.0375         & 0.0095             \\
Political Theory          & 0.0012         & .00001             \\
Eurozone                  & 0.0001         & .00002             \\
Taxation                  & 0.0135         & 0.0084             \\
Diet and Nutrition             & 0.0804         & 0.0139             \\
Education                 & 0.0011         & 0.0012             \\
Global Warming            & 0.0019         & 0.0011             \\
Terrorism                 & 0.0022         & 0.0019             \\ \hline
\end{tabular}
\end{table}

It is beyond our scope to detail the dynamics of all 22 topics.
However, a close look on one topic, Sensational Crime, shows
the kind of information this analysis obtains.
The posts about Sensational Crime largely concerned four events: 
the shooting of Trayvon Martin in February, the Aurora movie theater
shooting in July, the Sikh Temple shooting in August, and the Sandy Hook
shooting in December\footnote{Trayvon Martin was a young African American man shot by George Zimmerman, in what he claimed to be an act of self defense, while Martin was walking in Zimmerman's neighborhood. The Aurora theater massacre was a mass shooting at a movie theater in Aurora, Colorado. The Sikh Temple shooting was a mass shooting at a Sikh temple in Wisconsin. The Sandy Hook massacre was a mass shooting at an elementary school in Connecticut.}. 

To illustrate how the salience of a token changes over time, we use a
weighted frequency proportion which is equal to: 
\begin{equation}
WF_{w \in k} = \frac{P(Z_{d_t} = k_t \,|\, w \in d)F(w \in k_t)}{\sum_{w^*\in V}P(Z_{d_t} = k_t \,|\, w^* \in d)F(w^* \in k_t)},
\end{equation}
where $F(w \in k_t )$ is the frequency of the token $w$ in the $k$th
topic's distribution at time $t$. This weighted frequency can be
interpreted as the proportion of topic specific tokens at time $t$
that is taken up by token $w$, and is useful in this context as many
of the tokens are shared at high frequency between topics (such as
"people") and are therefore uninformative. So this quantity tracks the
topic specific information of a token over time. Recall that the topic-specific token-distributions are computed over the past 62 days, which accounts
for the smoothness of the curves. The gray shading around each curve
represents the 95\% Bayesian credible interval. 

\begin{figure}[H]
\includegraphics[width=.99\textwidth]{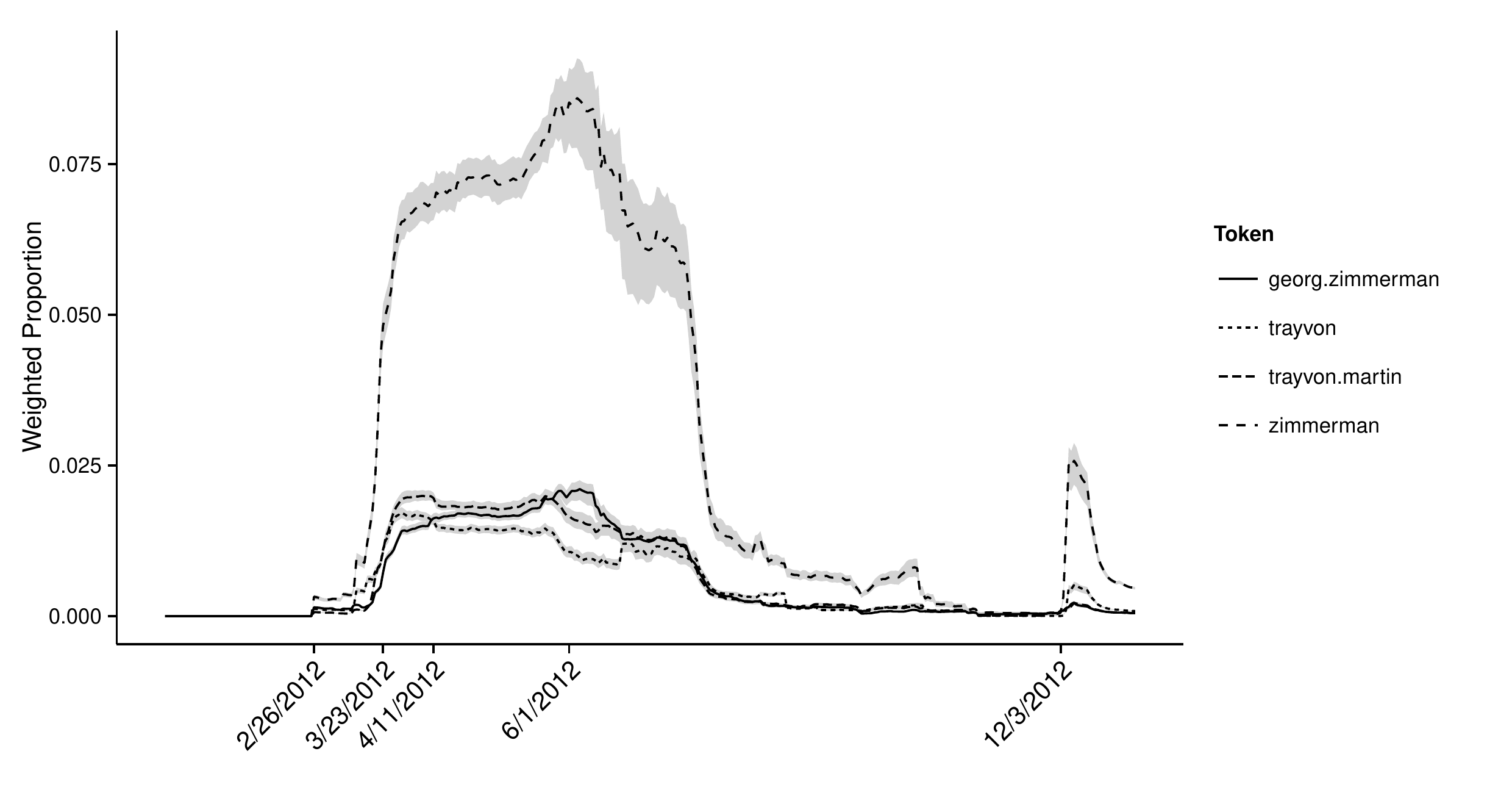}
\centering
\caption{Weighted proportion timeline of the Trayvon Martin shooting
and the subsequent legal case.  
The date 2/26/2012 is when Trayvon Martin was shot, 3/23/2012 is when
President Obama said that Trayvon could have been his son, 6/1/2012
was when Zimmerman's bond was revoked, and 12/3/2012 was when  
photos were released showing Zimmerman's injuries on the night of the
shooting. (95\% credible intervals shown.) } 
\label{zim}
\end{figure}

Figure \ref{zim}  presents the weighted frequency curves for tokens
specifically related to the shooting of Trayvon Martin. The plot shows
three interesting features. First, the prevalence of all of the tokens
does not spike up at the time of the shooting (2/26/2012), but rather
at the time of Obama's press statement regarding the shooting. Second,
the term ``zimmerman'' dominates the tokens, and in fact is the most
prevalent token in the whole of the Sensational Crime topic from March
22 to July 16th. The gap in prevalence between the tokens of
``zimmerman'' and ``trayvon'' or ``trayvon martin'' is also interesting,
suggesting that in this case, media attention was on the perpetrator
rather than the victim.  

\begin{figure}[H]
\includegraphics[width=.99\textwidth]{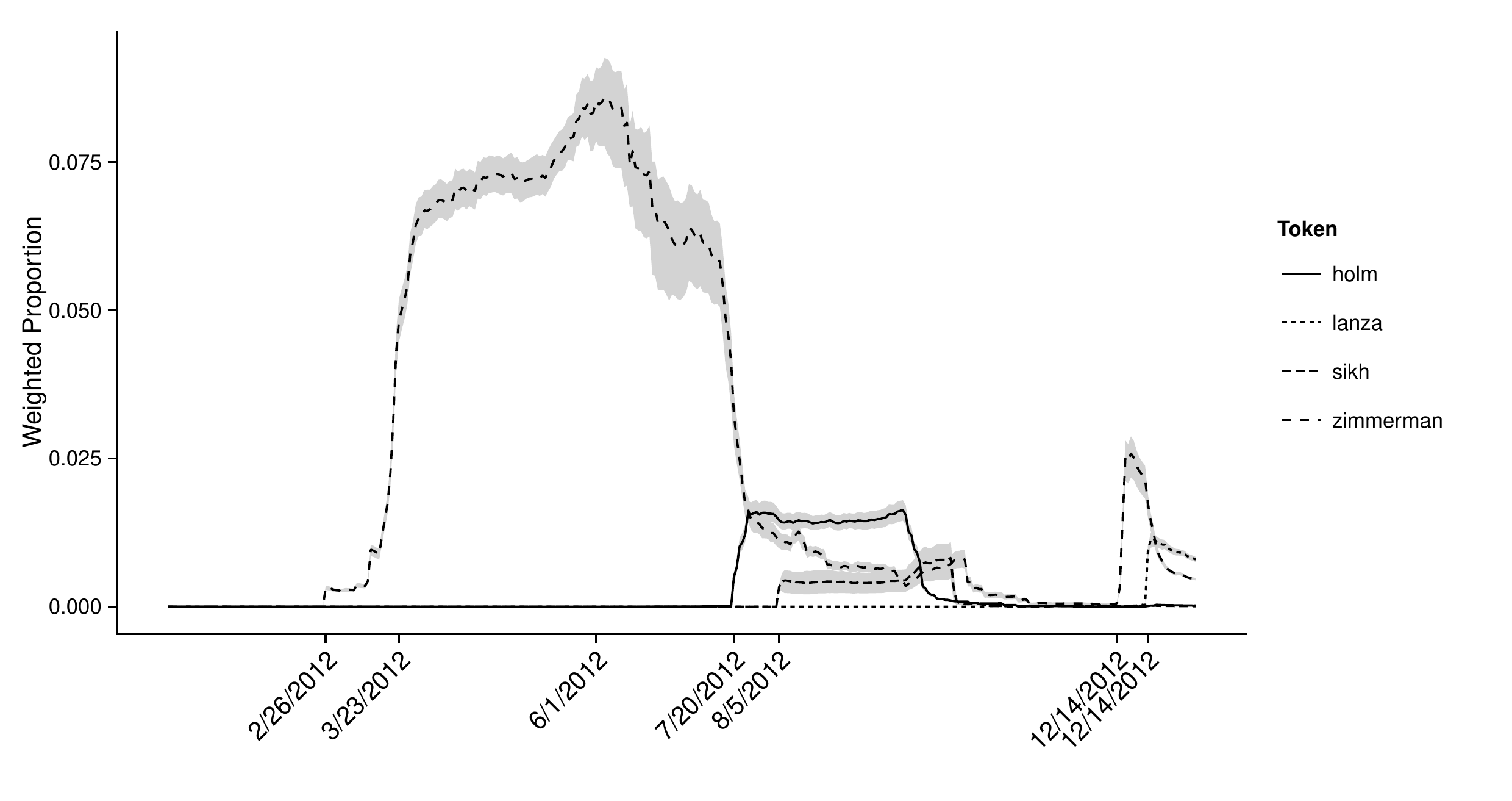}
\centering
\caption{Weighted proportion timeline of major events in the Sensational
Crime topic.   
The date 2/26/2012 is when Trayvon Martin was shot, 3/23/2012 is when
President Obama says that Trayvon could have been his son, 6/1/2012 is
when Zimmerman's bond was revoked, 7/20/2012 is the Aurora mass
shooting by James Holmes, 8/5/2012 is the Sikh Temple shooting by
Michael Page, 12/3/2012 is when photos showing Zimmerman's physical
injuries were released, and 12/14/2012 is when the Sandy Hook massacre  
occurred. (95\% credible intervals shown.)}
\label{year}
\end{figure}

Figure \ref{year} tracks the major events in the
Sensational Crime topic for the entire year. 
Notably, media focus is
never as strong on the tokens related to the events as it is on
``Zimmerman'' specifically. 
Rather, the usual top terms over the course of
the year are ``police'' and ``gun''. 
Also notable is the lack of events in
the later part of the year after the media attention on the Sikh
Temple Shooting receded. 

\subsection{The Network Results}

\deleted[id=TRH]{The use of network data improves the topic discovery, based on
a comparison to analysis of the same dataset by cite{pospisil}
that used only the text.
However, this analysis also enables us to use the topic modeling to 
improve community detection, through the incorporation of the
interest-group block structure.}

Table \ref{nettab} shows the posterior means of the network 
parameters with 95\% credible intervals. 
The edge parameter posterior mean indicates that the network is 
rather sparse at most time points. 
Interestingly, the 7-day lag parameter was negative, suggesting that 
blogs which were recently linked were less likely to link in the
near future.
There are two plausible explanations for this finding. 
First, the linkage dynamics may not be driven by recent links, 
but rather the links are a consequence of the events taking place. 
An upsurge in linking when an event occurs is followed by a decrease 
in the number of links as the event fades out of the news cycle. 
Second, if linking is done as part of a debate, then once a point
has been made, the bloggers may not feel a need for back-and-forth
argument.

The block parameter is strongly positive (mean = 1.058, standard
deviation 0.240), suggesting that blogs which share common interests are 
more likely to link to each other. 
This is particularly important, as the block statistic
was not only formed from explicit block matching, but also 
from blogs that did not share the same interests.
The block statistic is proportional to the shared topic interests. 
This result directly links the network model to the topic model, 
and allows the analyst to make claims about the block structure as 
inferred from the topics.  

Finally, and predictably, both the in-degree and out-degree of
a blog increases the probability that the block will receive
links. 
These parameters were included in the analysis to control for
the influence of highly popular blogs such as {\em The Blaze} and 
{\em The Huffington Post}.
  
\begin{table}[H]
\centering
\caption{Posterior means and 95\% credible intervals for network parameters.}
\label{nettab}
\begin{tabular}{lll}
\hline
Parameter             & Posterior Mean & 95\% CI              \\ \hline
Edges                 & -8.524         & {[}-8.539, -8.513{]} \\
7 day lag             & -0.163         & {[}-0.198, -0.131{]} \\
Block                 & 1.058          & {[}0.638, 1.485{]}   \\
Outdegree of Receiver & 0.330          & {[}0.329, 0.332{]}   \\
Indegree of Receiver  & 0.497          & {[}0.496, 0.499{]}   \\ \hline
\end{tabular}
\end{table}  

We can examine the link dynamics within a topic block. 
There were 21 blogs whose maximum posterior probability
of block assignment placed them in the block that was only
interested in the Sensational Crime topic. 
Only 2 of these 21 blogs received any
links over the course of the year, and only 1 received links
within the block (\textit{legalinsurrection.com}). 
While this runs counter to the idea that they form one block, 
recall that blogs are also more
likely to link to blogs that share some of the same topic
interest. 
There are a total of 62 blogs to which members of the Sensational Crime
block link, and 15 of these blogs receive approximately
90\% of the links. 
As such, the Sensational Crime topic block appears to be a set 
of ``commenter'' blogs that react to posts that are posted on 
larger blogs. 
Our model allows the analyst to isolate the blogs that post on 
a particular topic, to get a better idea of the linkage dynamics 
around important events. 
As an example, we describe how the
linkage pattern changes around the occurrence of Barack Obama's speech
regarding the shooting of Trayvon Martin, and also following the Aurora
shooting. 

\begin{figure}[H]
\includegraphics[width=.65\textwidth]{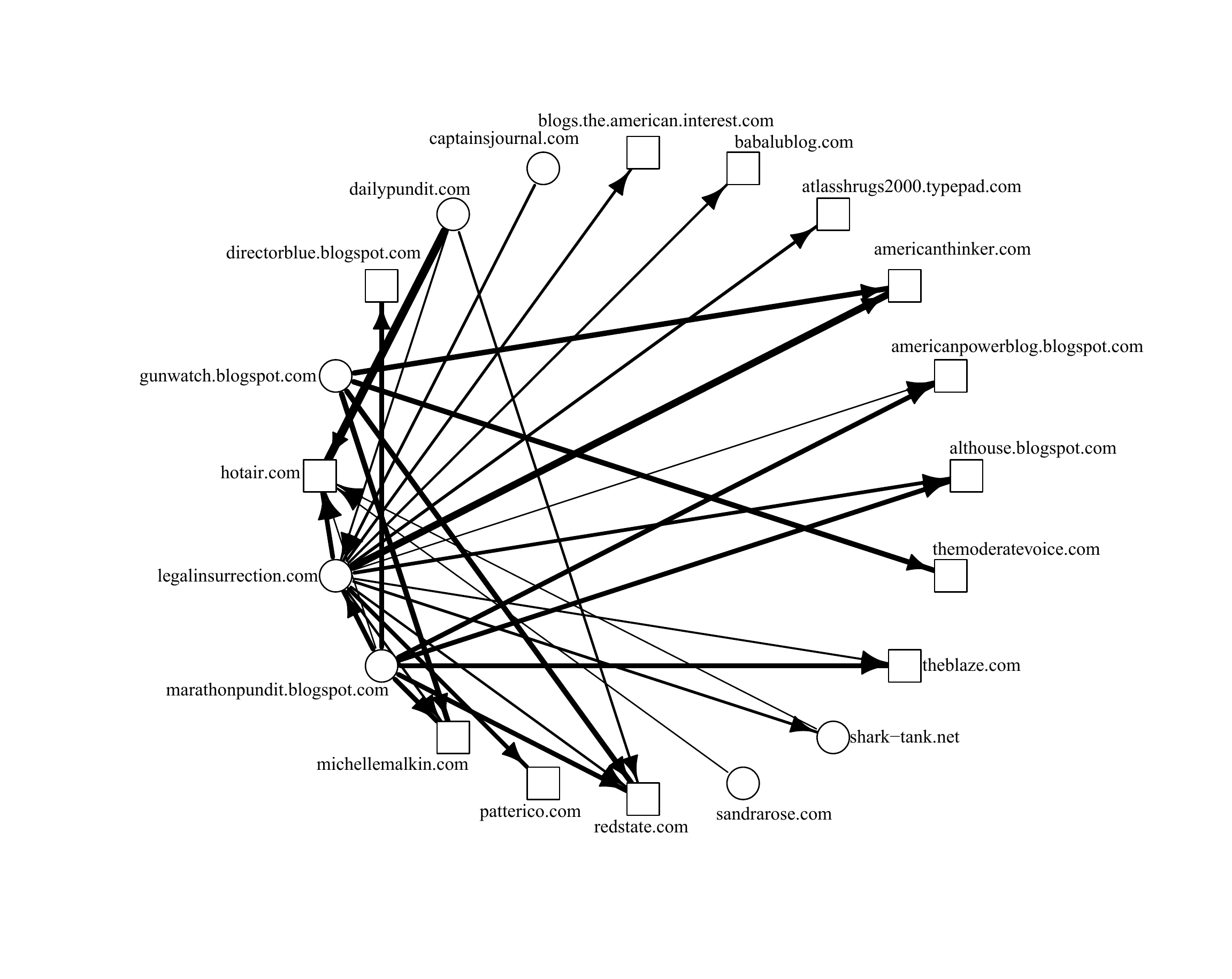}
\centering
\caption{Fifteen day aggregate linkage from 3/8/2012 to 3/22/2012,
immediately before President Obama's comment. 
The number of links, represented by line thickness, 
is root transformed for clarity. 
Circular nodes are blogs in the Sensational Crime block. 
Square nodes are blogs to which the Sensational Crime blogs link,
and these blogs are generally in multi-topic blocks, where one
of the topics is Sensational Crime.}
\label{T1}
\end{figure}

\begin{figure}[H]
\includegraphics[width=.65\textwidth]{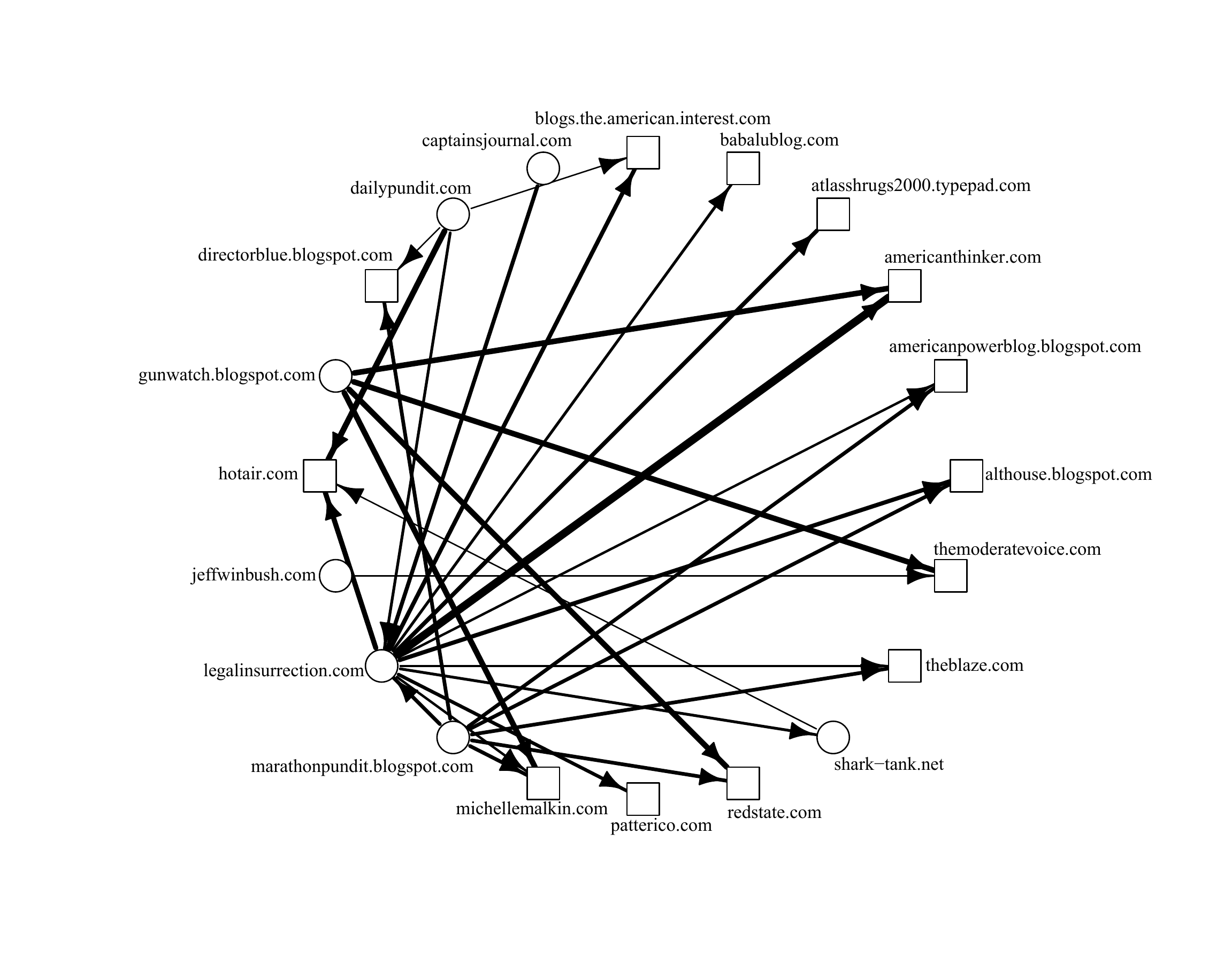}
\centering
\caption{This figure is constructed in the same way as
Fig.\ \ref{T1}, but for the time period from 3/23/2012 to
  4/6/2012, immediately after President Obama's comment.} 
\label{T2}
\end{figure}

\begin{figure}[H]
\includegraphics[width=.65\textwidth]{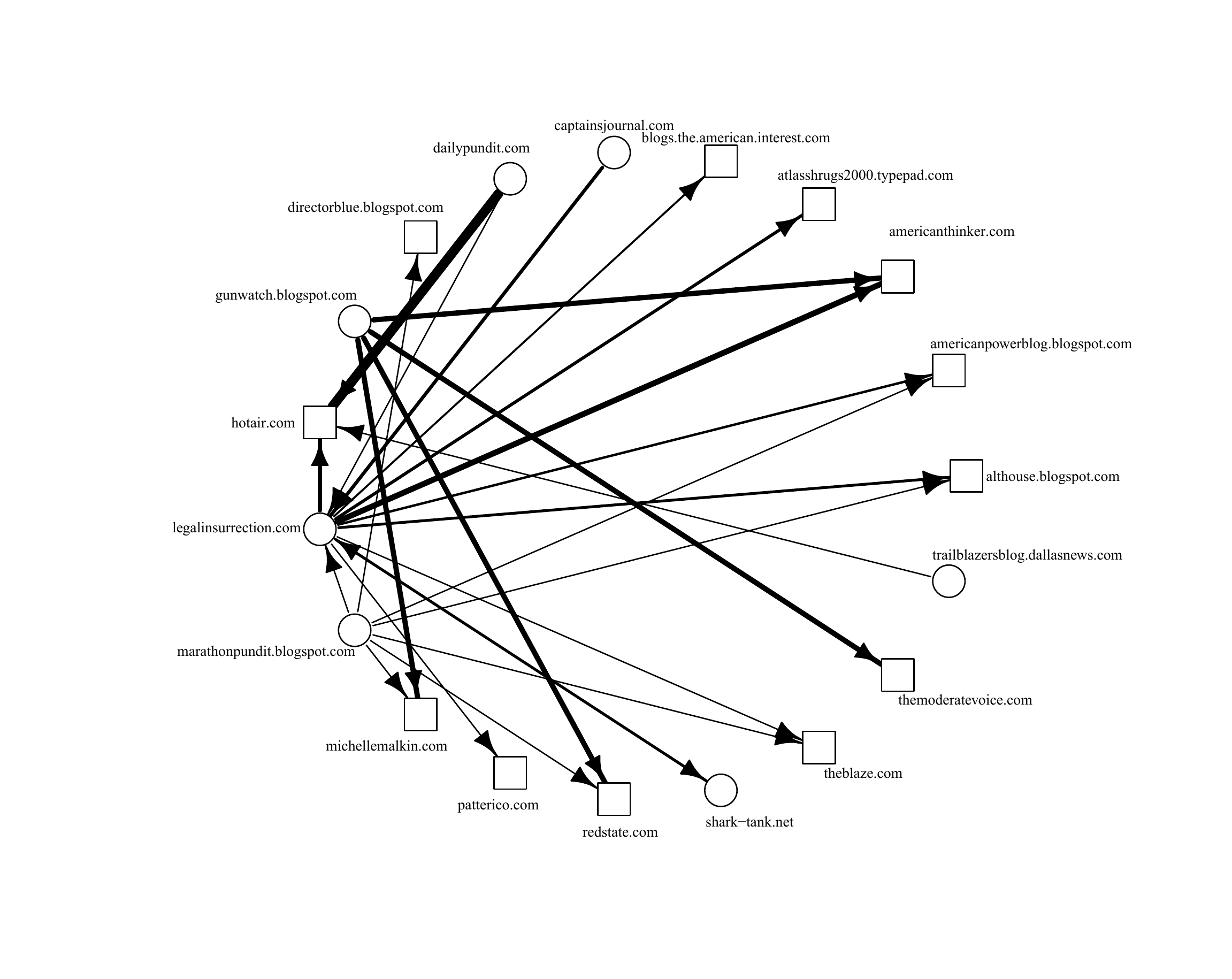}
\centering
\caption{This figure is constructed in the same way as
Fig.\ \ref{T1}, but for the time period from 7/5/2012 to
  7/19/2012, immediately before the Aurora shooting.} 
\label{A1}
\end{figure}

\begin{figure}[H]
\includegraphics[width=.65\textwidth]{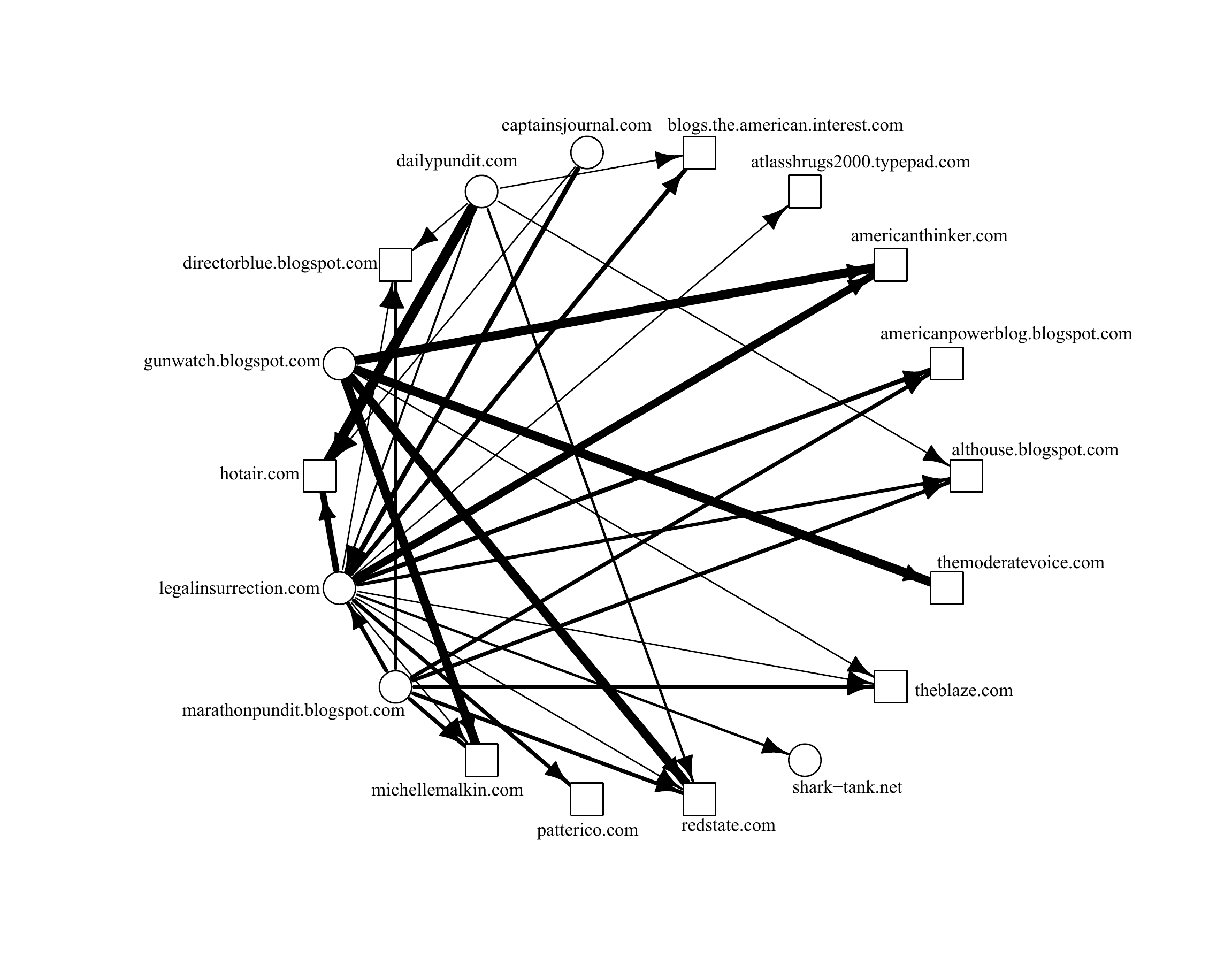}
\centering
\caption{This figure is constructed in the same way as
Fig.\ \ref{T1}, but for the time period from 7/20/2012 to
  8/2/2012, immediately after the Aurora shooting.} 
\label{A2}
\end{figure}

Figures \ref{T1} and \ref{T2} show the link structure from the blogs
in the Sensational Crime block to other blogs.
The data are aggregated over fifteen days.
Figure \ref{T1} pertains to the days before President Obama's
press conference regarding Trayvon Martin on 3/23/2013, and
Fig.\ \ref{T2} pertains the days following his remarks.
Figure \ref{A1} pertains to the period immediately
before the Aurora shooting on 7/20/2012, and Fig.~\ref{A2}
pertains to the period immediately after.

To improve interpretability, only a subset of blogs and links are
plotted. 
Specifically, blogs that were assigned to the block interested
only in Sensational Crime, and who posted during the specified time
frame, are plotted. 
Additionally, blogs who are part of the 15 blog subset that
received 90\% of the links from the Sensational Crime block,
and which received links within the timeframe, are
plotted. 
Also, links generated from blogs in the Sensational Crime
block to other members of the same block, or to
other blogs, are plotted. 
Links emanating from the 15-node subset are
not plotted. 
These plotting constraints help enable us to discern
and interpret the community structure that formed in
the discussion of these events.

Comparing Figs.\ \ref{T1} and \ref{T2} shows that the community
structure seen in the linkage pattern did not change much as a 
result of the press conference in which President Obama remarked
that if he had a son, he would resemble Trayvon Martin.
Remarkably, there was also no net increase in posting rates.
It is known that there was a flurry of posts at this time, and 
it turns out that uptick was allocated to the Election block,
as people speculated on how his remarks would affect the 
2012 presidential election. 

The patterns surrounding the Aurora shooting (Figs.\ref{A1} and
\ref{A2}) are more clear. 
The community structure in the discussion is essentially the same,
but the amount of traffic increases conspicuously.
Specifically, the number of links in the 15 days before the
shooting was 197, but afterwards it was 427.
Linkage rates especially increase from
\texttt{gunwatch.blogspot.com}. 
In general, this agrees with the conclusion that the methodology
is able to detect stable communities whose linkage rates are 
driven by news events.

To further illustrate the findings of the network model for a 
different block, we now present examples from the Election block. 
There are 52 blogs that the model assigned to the block whose
only interest was the presidential election.
Of these 52 blogs, 33 blogs linked to or received links from other 
blogs within this same block.
And of these 33 blogs, 12 were the recipients of all links. 
We use random walk community detection \citep{pons} upon the
Election block to show that the model can extract meaningful 
subnetworks for use in secondary analyses. 

\begin{figure}[H]
\includegraphics[width=.7\textwidth]{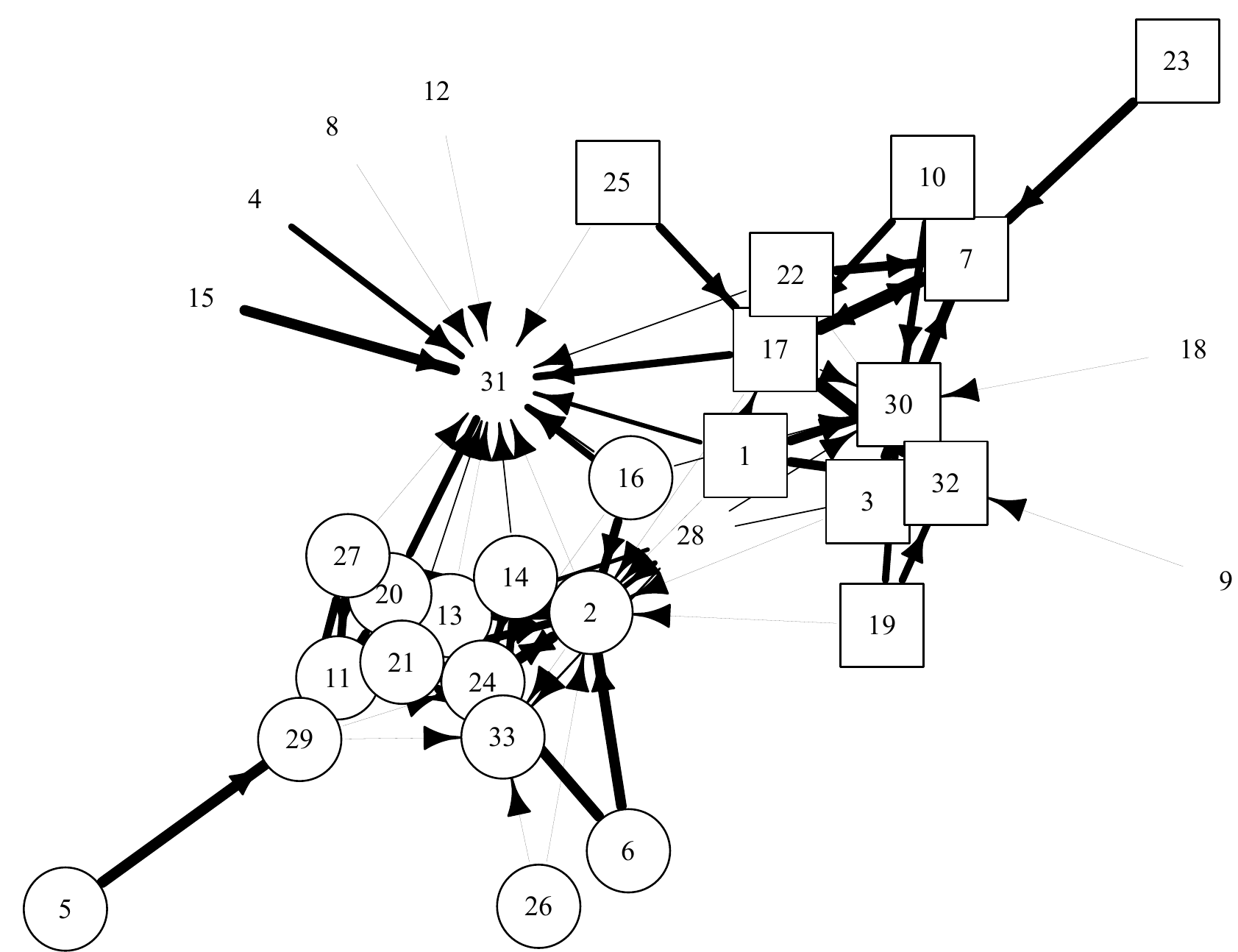}
\centering
\caption{Community substructure of the Election block. 
Circles and squares denote separate communities. 
The absence of shape denotes membership in a small community. 
The thickness of edges correspond to a log transformation of 
the number of links sent over the entire year.}
\label{E1}
\end{figure}

Figure \ref{E1}  shows the community substructure for the Election 
block aggregated over the entire year. 
Random walk community detection indicates that seven communities 
optimized modularity, but two communities contained the majority 
of the blogs. 
As such, for interpretability, only these two 
communities are shown. 
The modularity of this partition is 0.49, and a 10,000 sample
permutation test of the community labels indicated that this value of
modularity was in the 99th percentile (the greatest modularity found
in the permutation test was 0.317). 
This result indicates that the model found meaningful community
structure, rather than sampling
variability.  

Table \ref{my-label} contains the blog labels for the 
Election block network. 
Examination of the block structure shows that the majority of the
blogs partitioned into one of two communities. 
Based on \texttt{Technorati} ratings, the community plotted as 
circles in Figure~\ref{E1} is
politically conservative, while the other community plotted as
squares is liberal.
This separation of the two ends of the political spectrum has been 
found before in blogs \citep{Lawrence2010}. 
There is little communication between the two communities, 
but a lot of communication within those communities.
Interestingly, both communities sent many links to blog 31, which 
was allocated into a distinct community that it shared with 
blogs 15, 4 and 28. 
Blog 31 is
\textit{mediaite.com}, a non-partisan general news and media blog, and
the pattern of links from both partisan communities suggests that
\textit{mediaite.com} acts as a common source of information.

\begin{table}[H]
\small
\centering
\caption{Blog names and their community membership.}
\label{my-label}
\begin{tabular}{llc}
\hline
Label & Blog                               & Community \\ \hline
1     & afeatheradrift.wordpress.com       & 1         \\
2     & atlasshrugs2000.typepad.com        & 3         \\
3     & bleedingheartlibertarians.com      & 1         \\
4     & brainsandeggs.blogspot.com         & 2         \\
5     & citizentom.com                     & 3         \\
6     & crethiplethi.com                   & 3         \\
7     & crookedtimber.org                  & 1         \\
8     & davedubya.com                      & 4         \\
9     & dogwalkmusings.blogspot.com        & 5         \\
10    & driftglass.blogspot.com            & 1         \\
11    & greatsatansgirlfriend.blogspot.com & 3         \\
12    & hennessysview.com                  & 6         \\
13    & joshuapundit.blogspot.com          & 3         \\
14    & marezilla.com                      & 3         \\
15    & mediabistro.com                    & 2         \\
16    & michellesmirror.com                & 3         \\
17    & nomoremister.blogspot.com          & 1         \\
18    & ochairball.blogspot.com            & 7         \\
19    & patriotboy.blogspot.com            & 1         \\
20    & righttruth.typepad.com             & 3         \\
21    & rightwingnews.com                  & 3         \\
22    & rogerailes.blogspot.com            & 1         \\
23    & rwcg.wordpress.com                 & 1         \\
24    & sultanknish.blogspot.com           & 3         \\
25    & tbogg.firedoglake.com              & 1         \\
26    & thecitysquare.blogspot.com         & 3         \\
27    & therightplanet.com                 & 3         \\
28    & thoughtsandrantings.com            & 2         \\
29    & varight.com                        & 3         \\
30    & blogs.suntimes.com                 & 1         \\
31    & mediaite.com                       & 2         \\
32    & rightwingwatch.org                 & 1         \\
33    & patdollard.com                     & 3         \\ \hline
\end{tabular}
\end{table}

\section{Conclusion}

In this manuscript we present a \deleted[id=TRH]{novel}Bayesian model for analyzing \replaced[id=TRH]{a large dataset of political blog posts.}
{dynamic text networks.} 
This model links the network
dynamics to topic dynamics through a block structure that
informs both the topic assignment of a \replaced[id=TRH]{post}{document} and the
linkage pattern of the network. 

A major feature of our model is that the block structure enables 
interpretable associations among topics.
For example, there is a two-topic block
whose members are interested in both Election topic and the
Republican Primary topic, but there is no block whose members
are interested in just the Supreme Court and Global Warming.
That pattern of shared interest conforms to what one would
expect.

Another feature of our model is the flexibility of the network model. This analysis uses a limited set of predictors, but
the ERGM modeling framework can easily incorporate additional
covariates \citep{Robins2001b} and structural features. 
Additionally, if one uses a maximum
pseudo-likelihood approach \citep{frank} as a way of 
approximating the likelihood, then higher order subgraph terms, 
such as number of triangles or geometrically weighted edgewise shared
partners \citep{Hunter2008} can account for transitivity effects. 
Finally, while the block structure modeled in this paper was based
upon similarity in topic interest, more nuanced models are possible,
and these could use information on, say, political ideology, which
the analysis of the Election block found to be important in
predicting linkage patterns.

Finally, one major advantage of our approach to modeling this data is the nonparametric nature of
the topic dynamics. 
By avoiding an autoregressive specification of
topic dynamics, as in \citet{blei2006dynamic}, 
topics are able to change more freely; in particular,
it is possible for new tokens with high probability to emerge
overnight.
This is ideal for the blog data, since the blogs are often
responding to news events.

Our analysis of the political blog dataset had an interpretable
topic and block set, and analysis of the Sensational Crime block and
the Election block reached reasonable conclusions.
Specifically, the dominance of the token ``zimmerman'' across the
year agrees with our sense of the tone and primacy of that 
discussion, and the spike following the Aurora shooting is 
commensurate with its news coverage.
The Election block neatly split into subcommunities along
partisan lines, which accords with previous research
\citep{Lawrence2010}. 

\deleted[id=TRH]{Finally, the focus in this manuscript was the methodological
development, but there are many political science questions that
await deeper exploration.
Among these are study of the matrix that models topic activation,
which can be linked to specific headline events, the use of 
political affiliation as a covariate in the linkage dynamics, 
and examination of the association structure among blocks that
are interested in more than one topic.}

\added[id=TRH]{While our focus was on the specific application of the political blog data, the model developed here has features that can generalize to other dynamic text networks.}
such as the Wikipedia and scientific citation networks. \added[id=TRH]{Specifically, the connection of topic and link structure through a block structure allows for document content to inform the community structure of the overall network.}
However, each application requires some hand fitting that
captures specific aspects of the data.
For example, the block structure might need to be dynamic; this
would make sense is scientific citation networks, since disciplines
sometimes bifurcate (e.g., the computer science of 1970 has now split
into machine learning, quantum computing, algorithms, and many other
subfields).
Also, scientific citation is strictly directional in time---one cannot
cite future articles.
But the Wikipedia is not directional in time; an article posted a year
ago can send a link to one posted yesterday.
So specific applications will require tinkering with the model
described here.

\added[id=TRH]{The work presented here suggests several avenues of future research. On the methodological side, the model can be generalized and extended in several ways. Specifically, the block membership could be considered a dynamic property, allowing blogs to change interest in topics over time. Additionally, building a dynamic model for link patterns would allow researchers to examine specific properties of links over time, rather than assuming the same link generating distribution at all time points. Finally, this model can be adapted to other dynamic text networks, and its performance should be compared to more traditional topic analysis and community detection procedures.}




\vfill

\bibliography{mcs.bib}

\begin{thebibliography}{}

\bibitem[Airoldi et~al., 2008]{Airoldi2008b}
\MakeUppercase{Airoldi, E.~M.}, \MakeUppercase{Blei, D.~M.},
  \MakeUppercase{Fienberg, S.~E.}, and \MakeUppercase{Xing, E.~P.} (2008).
\newblock {Mixed Membership Stochastic Blockmodels}.
\newblock {\em Journal of Machine Learning Research}, 9(2008):1981--2014.

\bibitem[Arun et~al., 2010]{arun}
\MakeUppercase{Arun, R.}, \MakeUppercase{Suresh, V.}, \MakeUppercase{Madhavan,
  C.~V.}, and \MakeUppercase{Murthy, M.~N.} (2010).
\newblock On finding the natural number of topics with latent dirichlet
  allocation: Some observations.
\newblock In {\em Advances in Knowledge Discovery and Data Mining}, pages
  391--402. Springer.

\bibitem[Blei et~al., 2003]{Blei2003}
\MakeUppercase{Blei, D.}, \MakeUppercase{Ng, A.}, and \MakeUppercase{{and M.
  Jordan}} (2003).
\newblock {Latent Dirichlet Allocation}.
\newblock {\em Journal of Machine Learning Research}, 3:993--1022.

\bibitem[Blei and Lafferty, 2006]{blei2006dynamic}
\MakeUppercase{Blei, D.~M.} and \MakeUppercase{Lafferty, J.~D.} (2006).
\newblock Dynamic topic models.
\newblock In {\em Proceedings of the 23rd international conference on Machine
  learning}, pages 113--120. ACM.

\bibitem[Brown et~al., 1992]{brown}
\MakeUppercase{Brown, P.~F.}, \MakeUppercase{Desouza, P.~V.},
  \MakeUppercase{Mercer, R.~L.}, \MakeUppercase{Pietra, V. J.~D.}, and
  \MakeUppercase{Lai, J.~C.} (1992).
\newblock Class-based n-gram models of natural language.
\newblock {\em Computational linguistics}, 18(4):467--479.

\bibitem[Chang and Blei, 2009]{chang}
\MakeUppercase{Chang, J.} and \MakeUppercase{Blei, D.~M.} (2009).
\newblock Relational topic models for document networks.
\newblock In {\em International conference on artificial intelligence and
  statistics}, pages 81--88.

\bibitem[Faust and Wasserman, 1992]{faust}
\MakeUppercase{Faust, K.} and \MakeUppercase{Wasserman, S.} (1992).
\newblock Blockmodels: Interpretation and evaluation.
\newblock {\em Social networks}, 14(1):5--61.

\bibitem[Frank and Strauss, 1986]{frank}
\MakeUppercase{Frank, O.} and \MakeUppercase{Strauss, D.} (1986).
\newblock Markov graphs.
\newblock {\em Journal of the American Statistical Association},
  81(395):832--842.

\bibitem[Gilks et~al., 1995]{gilks}
\MakeUppercase{Gilks, W.~R.}, \MakeUppercase{Best, N.}, and \MakeUppercase{Tan,
  K.} (1995).
\newblock Adaptive rejection metropolis sampling within gibbs sampling.
\newblock {\em Applied Statistics}, pages 455--472.

\bibitem[Ho et~al., 2012]{ho}
\MakeUppercase{Ho, Q.}, \MakeUppercase{Eisenstein, J.}, and
  \MakeUppercase{Xing, E.~P.} (2012).
\newblock Document hierarchies from text and links.
\newblock In {\em Proceedings of the 21st international conference on World
  Wide Web}, pages 739--748. ACM.

\bibitem[Hoff et~al., 2002]{Hoff2002}
\MakeUppercase{Hoff, P.~D.}, \MakeUppercase{Raftery, A.~E.}, and
  \MakeUppercase{Handcock, M.~S.} (2002).
\newblock {Latent Space Approaches to Social Network Analysis}.
\newblock {\em Journal of the American Statistical Association},
  97(460):1090--1098.

\bibitem[Hoffman et~al., 2010]{hoffman2010online}
\MakeUppercase{Hoffman, M.}, \MakeUppercase{Bach, F.~R.}, and
  \MakeUppercase{Blei, D.~M.} (2010).
\newblock Online learning for latent dirichlet allocation.
\newblock In {\em Advances in Neural Information Processing Systems}, pages
  856--864.

\bibitem[Holland and Leinhardt, 1981]{holland}
\MakeUppercase{Holland, P.~W.} and \MakeUppercase{Leinhardt, S.} (1981).
\newblock An exponential family of probability distributions for directed
  graphs.
\newblock {\em Journal of the american Statistical association},
  76(373):33--50.

\bibitem[Hubert and Arabie, 1985]{hubert}
\MakeUppercase{Hubert, L.} and \MakeUppercase{Arabie, P.} (1985).
\newblock {Comparing partitions}.
\newblock {\em Journal of Classification}, 2(1):193--218.

\bibitem[Hunter et~al., 2008]{Hunter2008}
\MakeUppercase{Hunter, D.~R.}, \MakeUppercase{Goodreau, S.~M.}, and
  \MakeUppercase{Handcock, M.~S.} (2008).
\newblock {Goodness of Fit of Social Network Models}.
\newblock {\em Journal of the American Statistical Association},
  103(481):248--258.

\bibitem[Krivitsky and Handcock, 2014]{krivitsky}
\MakeUppercase{Krivitsky, P.~N.} and \MakeUppercase{Handcock, M.~S.} (2014).
\newblock A separable model for dynamic networks.
\newblock {\em Journal of the Royal Statistical Society: Series B (Statistical
  Methodology)}, 76(1):29--46.

\bibitem[Latouche et~al., 2011]{latouche}
\MakeUppercase{Latouche, P.}, \MakeUppercase{Birmel{\'{e}}, E.}, and
  \MakeUppercase{Ambroise, C.} (2011).
\newblock {Overlapping stochastic block models with application to the French
  political blogosphere}.
\newblock {\em Annals of Applied Statistics}, 5(1):309--336.

\bibitem[Lawrence et~al., 2010]{Lawrence2010}
\MakeUppercase{Lawrence, E.}, \MakeUppercase{Sides, J.}, and
  \MakeUppercase{Farrell, H.} (2010).
\newblock {Self-Segregation or Deliberation? Blog Readership, Participation,
  and Polarization in American Politics}.
\newblock {\em Perspectives on Politics}, 8(01):141.

\bibitem[McNamee and Mayfield, 2003]{mcnamee}
\MakeUppercase{McNamee, P.} and \MakeUppercase{Mayfield, J.} (2003).
\newblock Jhu/apl experiments in tokenization and non-word translation.
\newblock In {\em Comparative Evaluation of Multilingual Information Access
  Systems}, pages 85--97. Springer.

\bibitem[Moody, 2004]{moody}
\MakeUppercase{Moody, J.} (2004).
\newblock The structure of a social science collaboration network: Disciplinary
  cohesion from 1963 to 1999.
\newblock {\em American Sociological Review}, 69(2):213--238.

\bibitem[Newman and Girvan, 2004]{Newman2004a}
\MakeUppercase{Newman, M. E.~J.} and \MakeUppercase{Girvan, M.} (2004).
\newblock {Finding and evaluating community structure in networks}.
\newblock {\em Physical Review E - Statistical, Nonlinear, and Soft Matter
  Physics}, 69(2 2):026113.

\bibitem[Pons and Latapy, 2006]{pons}
\MakeUppercase{Pons, P.} and \MakeUppercase{Latapy, M.} (2006).
\newblock {Computing communities in large networks using random walks}.
\newblock {\em J. Graph Algorithms Appl.}, 10(2):191.

\bibitem[Ramos, 2003]{ramos}
\MakeUppercase{Ramos, J.} (2003).
\newblock Using tf-idf to determine word relevance in document queries.
\newblock In {\em Proceedings of the first instructional conference on machine
  learning}.

\bibitem[Robins et~al., 2001]{Robins2001b}
\MakeUppercase{Robins, G.}, \MakeUppercase{Elliott, P.}, and
  \MakeUppercase{Pattison, P.} (2001).
\newblock {Network models for social selection processes}.
\newblock {\em Social Networks}, 23(1):1--30.

\bibitem[Snijders and Nowicki, 1997]{snijders97}
\MakeUppercase{Snijders, T.~A.} and \MakeUppercase{Nowicki, K.} (1997).
\newblock Estimation and prediction for stochastic blockmodels for graphs with
  latent block structure.
\newblock {\em Journal of Classification}, 14(1):75--100.

\bibitem[Steinley, 2004]{steinley}
\MakeUppercase{Steinley, D.} (2004).
\newblock Properties of the hubert-arable adjusted rand index.
\newblock {\em Psychological methods}, 9(3):386.

\bibitem[Technorati, 2002]{tech}
\MakeUppercase{Technorati} (2002).
\newblock {https://web.archive.org/web/20140420052710/http://technorati.com/}.

\bibitem[Wang et~al., 2011]{WangD}
\MakeUppercase{Wang, E.}, \MakeUppercase{Silva, J.}, \MakeUppercase{Willett,
  R.}, and \MakeUppercase{Carin, L.} (2011).
\newblock Dynamic relational topic model for social network analysis with noisy
  links.
\newblock In {\em Statistical Signal Processing Workshop (SSP), 2011 IEEE},
  pages 497--500. IEEE.

\bibitem[Wasserman and Pattison, 1996]{Wasserman1996a}
\MakeUppercase{Wasserman, S.} and \MakeUppercase{Pattison, P.} (1996).
\newblock {Logit models and logistic regressions for social networks: I. An
  introduction to Markov graphs and $p^*$}.
\newblock {\em Psychometrika}, 61(3):401--425.

\bibitem[Yin and Wang, 2014]{yin}
\MakeUppercase{Yin, J.} and \MakeUppercase{Wang, J.} (2014).
\newblock A dirichlet multinomial mixture model-based approach for short text
  clustering.
\newblock In {\em Proceedings of the 20th ACM SIGKDD international conference
  on Knowledge discovery and data mining}, pages 233--242. ACM.

\end{thebibliography}



\end{document}